\documentclass[a4paper,10pt]{article}
\pdfoutput=1 % if your are submitting a pdflatex (i.e. if you have
\usepackage{jheppub} % for details on the use of the package, please
\usepackage[utf8]{inputenc}
\usepackage[T1]{fontenc}
\usepackage[english]{babel}
\usepackage{fancyhdr}
\usepackage{booktabs}
\usepackage{hyperref}
\usepackage{amssymb,amsfonts,amsmath}
\usepackage{graphics,graphicx,pstricks,color,colortbl}
\usepackage{slashed}
\usepackage{cleveref}
\crefname{figure}{Figure}{Figures}
\usepackage{subcaption}
\usepackage{multirow}
\usepackage{booktabs}
\usepackage{axodraw2}

\title{\boldmath Phase transitions and gravitational waves in a non-abelian vector dark matter scenario}

\author[a]{Nico Benincasa}

\author[b,c]{, Luigi Delle Rose}

\author[b,c]{, Luca Panizzi}

\author[b,c]{, Maimoona Razzaq}

\author[b]{, Savio Urzetta}

\affiliation[a]{School of Physics, University of Electronic Science and Technology of China, 611731 Chengdu, China}
\affiliation[b]{Dipartimento di Fisica, Università della Calabria, Arcavacata di Rende, I-87036, Cosenza, Italy}
\affiliation[c]{INFN, Gruppo Collegato di Cosenza, Arcavacata di Rende, I-87036, Cosenza, Italy}

\emailAdd{nico.benincasa@kbfi.ee}
\emailAdd{luigi.dellerose@unical.it}
\emailAdd{luca.panizzi@unical.it}
\emailAdd{maimoona.razzaq@unical.it}
\emailAdd{sav.urze98@gmail.com}

\abstract{We study a scenario where the Standard Model is extended by a SU(2) gauge group in the dark sector. The three associated dark gauge bosons are stabilised via a custodial symmetry triggered by an additional dark SU(2) scalar doublet, thus making them viable dark-matter candidates. After considering the most recent constraints for this model, we analyse the phase transition dynamics and compute the power spectrum of resulting stochastic gravitational-wave background. Finally, we find regions of the parameter space yielding the observed dark-matter relic density while also leading to strong enough phase transition with an associated gravitational-wave signal reaching the sensitivity of future space-based gravitational-wave detector, such as LISA, DECIGO, BBO, TianQin or Taiji.
}

\begin{document}

\maketitle

\section{Introduction}

The possibility to test models of new physics in complementary ways allows to effectively narrow down the possible theories which aim to solve the open problems of the Standard Model (SM) of particle physics. For many years, this complementarity involved a combination of different astrophysical and collider observations. For a decade, the observation of gravitational waves (GWs)~\cite{LIGOScientific:2016aoc, LIGOScientific:2017vwq} added a precious source of information for this kind of phenomenological approach. Indeed, scenarios of new physics usually predict new particles which, throughout the history of the Universe, might have contributed to phenomena leading to the generation of GWs, possibly in the range of observation of current or future experiments. In particular, it has been demonstrated that a first-order phase transition (FOPT) in the early Universe leads to a stochastic gravitational-wave background~\cite{Witten:1984rs, Hogan:1986dsh} which could be detected today by future space-based gravitational-wave observatories such as LISA~\cite{2017arXiv170200786A}, DECIGO~\cite{Kawamura:2011zz}, BBO~\cite{Corbin:2005ny}, TianQin~\cite{TianQin:2015yph}, or Taiji~\cite{Hu:2017mde}. Within the frame of the SM, the electroweak phase transition (PT) is a crossover~\cite{Kajantie:1996mn}, hence the necessity to add beyond-SM ingredients to make it first order. 

In this paper we consider a theoretical model which predicts dark matter (DM) candidates and at the same time can lead to FOPTs, allowing to probe its parameter space through GWs experiments. In this model, introduced in~\cite{Hambye:2008bq} and further studied in~\cite{Hambye:2009fg, Chen:2009ab, Diaz-Cruz:2010czr, Bhattacharya:2011tr, Hambye:2013dgv, Fraser:2014yga, Gross:2015cwa, Bernal:2015ova, DiChiara:2015bua, Barman:2017yzr, Baldes:2018emh, Barman:2019lvm, Hall:2019ank, Marfatia:2020bcs, Hisano:2020qkq, Baouche:2021wwa, Belyaev:2022shr, Abe:2023yte, Arteaga:2024vde, ArteagaTupia:2025awh}, the DM candidates appear in the form of massive gauge bosons of a non-Abelian $SU(2)$ group in the dark sector. Their stability is ensured by a custodial symmetry in the dark scalar sector which generates gauge boson masses through a mechanism of spontaneous symmetry breaking. The scalar doublet under the new $SU(2)$ can interact with the Higgs doublet via a quartic coupling, and the new physical scalar which is generated after the symmetry breaking mixes with the Higgs boson. In this way, the dark sector can communicate with the SM other than through gravity, potentially leading to observable signals.

Our analysis consists of the following steps. We will first identify the allowed regions of the four-dimensional parameter space of the model by combining theoretical constraints, observables from the Higgs sector of the SM and its precision tests in the electroweak sector, and multiple astrophysical constraints, updating previous analyses using the most recent experimental data. Subsequently, we will study the scalar sector of the model to determine the region of parameter space which can induce FOPTs during cosmological evolution, with the generation of GWs~\cite{Athron:2023xlk}. We will then evaluate whether the GWs can be observable by future experiments, and in which frequency range. 

This paper is structured as follows: in \cref{sec:model} we provide a short description of the vector DM model from a dark $SU(2)$ gauge group; in \cref{sec:DMbounds} the astrophysical and cosmological constraints are described and quantitatively discussed; in \cref{sec:thcollider} the theoretical and collider constraints are analysed; in \cref{sec:PTGW} the quantum and finite-temperature corrections to the scalar potential as well as the cosmic phase-transition parameters needed to compute the power spectrum of the subsequent generated stochastic gravitational-wave background are provided; in \cref{sec:allresults} we show the combination of all constraints and the allowed regions of parameter space, and determine the subregions where a first-order phase transition is possible, and finally compare the spectrum of predicted GWs with the reach of different future space-based gravitational-wave detectors.

\section{The $SU(2)_D$ model}
\label{sec:model}

For our study we consider a minimal SM extension, introduced in \cite{Hambye:2008bq}, providing non-abelian gauge boson DM candidates whose stability is guaranteed by a custodial symmetry built in the new scalar sector, for which we will describe here the main features. 

In this model, the SM gauge sector is extended with a new $SU(2)_D$ gauge group, where the $D$ suffix stands for ``dark'', under which all SM particles transform as singlets. The three gauge bosons $V_D^a$ acquire mass through a mechanism of spontaneous symmetry breaking, in which a new scalar boson $\Phi_D$, doublet under $SU(2)_D$, acquires a vacuum expectation value (VEV) $v_D$. The gauge bosons become then degenerate with mass $m_{V_D}=g_D v_D/2$, where $g_D$ is the coupling constant of $SU(2)_D$. The $\Phi_D$ is singlet under the SM gauge group, and therefore its only possible interaction with the SM is through a quartic portal coupling with the Higgs boson. The Lagrangian of the new-physics part of the model is
\begin{equation}
\label{eq:Lag}
\mathcal{L} = - \frac{1}{4} F_{D\mu\nu} F^{\mu\nu}_D + (D_\mu \Phi_D)^\dagger (D^\mu \Phi_D) + \mu_D^2 (\Phi_D^\dagger \Phi_D) - \lambda_D (\Phi_D^\dagger \Phi_D)^2 - \lambda_{HD} (\Phi_D^\dagger \Phi_D)(\Phi^\dagger \Phi)\;,
\end{equation}
where $F_D^{\mu\nu}$ is the $SU(2)_D$ field strength; the covariant derivative is $D_\mu=\partial_\mu - ig_D T_D^a V_D^a$, with $T_D$ the dark isospin. 
This Lagrangian is invariant under an $SO(4)$ symmetry, which is broken to the custodial $SO(3)$ symmetry by the VEV of $\Phi_D$, ensuring the stability of the $V_D^a$ gauge bosons, and making them viable dark matter candidates. An enlarged fermionic sector could break such symmetry, lift the degeneration between the dark vector bosons at loop level and allow for purely SM decay modes for one of the $V_D^a$~\cite{Belyaev:2022zjx,Belyaev:2022shr}.

Out of the 8 degrees of freedom in the scalar sector (the components of the Higgs and dark scalar doublets), 6 are Goldstone bosons which give mass to the electroweak gauge bosons and to the three $V_D^a$ of $SU(2)_D$. The remaining two correspond to physical states, the Higgs boson $h$ and a further CP-even scalar $H_D$. These states arise from the mixing between $\phi$ and $\phi_D$, appearing in the neutral components  of the Higgs and dark doublet respectively, $\Phi_{-1/2} = {1\over\sqrt2}( \phi+v)$ and $\Phi_{D,-1/2} = {1\over\sqrt2}( \phi_D + v_D)$ in unitary gauge, where $v\simeq 246$ GeV is the SM Higgs VEV. The scalar mass matrix in the vacuum $(v,v_D)$, 
$\left(
\begin{array}{cc}
\lambda_H v^2 & {\lambda_{HD}\over2} v v_D \\
{\lambda_{HD}\over2} v v_D & \lambda_D v_D^2
\end{array}
\right)$ is diagonalised by the rotation matrix $\left(\begin{array}{cc} \cos\theta_S & \sin\theta_S \\ -\sin\theta_S & \cos\theta_S \end{array}\right)$, where $\theta_S$ is the scalar mixing angle and $\lambda_H$ are the usual Higgs self-quartic coupling. 
In the absence of further new particles, mixing between EW and dark gauge sectors is forbidden because the scalar doublets transform under only one of the two gauge groups.

The full model contains four free parameters from the new sector. Following~\cite{Belyaev:2022shr}, we choose to identify them with the dark gauge coupling, the $V_D$ mass , the $H_D$ mass, and the scalar mixing angle
\begin{equation}
g_D,~m_{V_D},~m_{H_D},~\theta_S,\;
\end{equation}
which can be easily associated to physical observables.
The conversion map between these parameters and the Lagrangian ones is the following:
\begin{eqnarray}
\mu^2_H &=& {1\over2} \left(m_h^2\cos^2\theta_S +m_{H_D}^2\sin^2\theta_S +{1\over2} {g\over g_D} {m_{V_D}\over m_W} (m_{H_D}^2-m_h^2) \sin2\theta_S \right)\;, \\
\mu_D^2 &=& {1\over2} \left(m_h^2\sin^2\theta_S +m_{H_D}^2\cos^2\theta_S +{1\over2} {g\over g_D} {m_{V_D}\over m_W} (m_{H_D}^2-m_h^2) \sin2\theta_S \right)\;, \\
\lambda_H &=& {g^2\over 8m_W^2} \left(m_h^2\cos^2\theta_S +m_{H_D}^2\sin^2\theta_S\right)\;, \\
\lambda_D &=& {g_D^2\over 8m_{V_D}^2} \left(m_h^2\sin^2\theta_S +m_{H_D}^2\cos^2\theta_S\right)\;,\\
\lambda_{HD} &=& \frac{gg_{D}}{8m_{W}m_{V_D}}(m_{H_D}^2-m_{h}^2)\sin2\theta_S ,
\end{eqnarray}
with the Higgs mass parameter $\mu_H$, the Higgs mass $m_h\simeq 125$ GeV, the SU(2)$_L$ gauge coupling $g$ and the W$^\pm$ boson mass $m_W$.

\section{Testing the model}

This model is a minimal extension of the SM, yet it can be constrained from multiple different directions. In the following, we will first consider the theoretical bounds coming from ensuring that the scalar potential is bounded from below and from perturbative unitarity; then we will discuss cosmological and astrophysical bounds associated with the presence of a dark matter candidate; finally we will discuss how collider observables constrain modifications in the scalar sector.

From the numerical point of view, we have performed random scans on the four free parameters of the model: $m_{V_D}$ and $m_{H_D}$ logarithmically from 10 MeV to 100 TeV, $g_D$ also logarithmically from $10^{-5}$ to $4\pi$, and $\cos\theta_S$ linearly from 0 to 1. Given the vast range of the parameter space, we have also performed dedicated scans selecting only points satisfying one or all of the constraints discussed in the following sections, in order to increase the density of valid points for plotting purposes. Furthermore, we will be almost always showing results in the $\{m_{V_D},m_{H_D}\}$ plane, to allow for a direct comparison of various constraints in the same projection of the parameter space, and also to avoid presenting a too large number of figures, which might disrupt the flow of the discussion. Other projections for further investigation can be easily produced with the information provided below.
% In our analysis we did not take into account higher-order effects in the $g_D$ expansion. For this reason, when relevant, we will explicitely identify regions where points have $g_D$ larger than 1.

% \NB{also mention the refined scan made for cosmo only plot and the final one to focus on the underabundant region and have more statistics for the PT analysis}

% \NB{discard points with $g>1$} \LP{I think we can just comment. I wrote a tentative sentence}

\subsection{Cosmological and astrophysical bounds}
\label{sec:DMbounds}

\begin{figure}[h!]
\centering
\includegraphics[width=.9\textwidth]{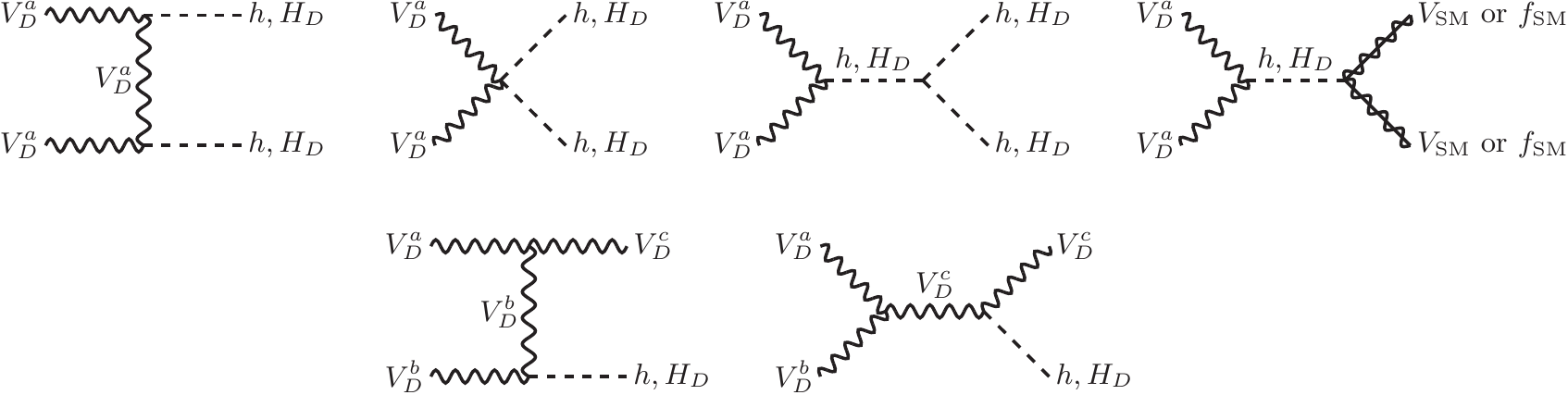}
\caption{\label{fig:annihilationtopologies} Annihilation processes, involving SM particles or $H_D$ in the final states (top row) or $V_D$ and scalar particles in the final state (bottom row). $V_\text{SM}$ and $f_\text{SM}$ respectively stand for massive vector and fermion SM particles.}
\end{figure}

The model parameters can be constrained by requiring that the relic density of the dark gauge bosons does not exceed the measured value by the Planck experiment~\cite{Planck:2018vyg}, with a precision smaller than 1\%:
\begin{equation}
\Omega_{\rm{DM}} h^2 = 0.120\pm0.001\;,
\end{equation}
with $h$ the dimensionless Hubble constant. The annihilation processes which determine the relic density of the DM candidates in this model are of two types: those involving exclusively SM particles and/or $H_D$ in the final state and those in which two $V_D^a$ annihilate into one $V_D^a$ and a scalar~\cite{Hambye:2008bq}. Representative annihilation processes of the two kinds are shown in~\cref{fig:annihilationtopologies}.

Furthermore, the model can be tested against experimental data from different direct detection (DD) experiments, considering that the DM states can interact with nuclei through the Higgs and $H_D$ boson. 

Both relic density and direct detection constraints have been computed using {\tt MicrOMEGAs v6}~\cite{Alguero:2023zol}, which allows testing the signal against data from multiple experiments, such as LZ~\cite{LZ:2022ufs}, PICO-60~\cite{PICO:2017tgi}, XENON1T~\cite{XENON:2018voc} and PandaX-4T~\cite{PandaX-4T:2021bab}, for DM masses ranging from few GeV to 10 TeV, or DarkSide-50~\cite{DarkSide:2018bpj} and CRESST-III~\cite{CRESST:2019jnq}. These experiments are sensitive to DM masses down to a few hundreds MeV, which is contained in the range of our scan. The annihilation of DM into $h$ and $H_D$ after they concentrate in the center of the Sun and the Earth also allows to set bounds from the neutrinos arising from the subsequent decays of the scalars. The predicted neutrino fluxes can be computed and tested against IceCube (IC) data through {\tt MicrOMEGAs}~\cite{Belanger:2015hra}.

Indirect detection (ID) constraints are evaluated through {\sc MadDM}~\cite{Ambrogi:2018jqj} using data from gamma-ray lines and gamma-ray continuum from the Fermi-LAT analysis of dwarf spheroidal galaxies~\cite{Fermi-LAT:2016uux}
% we did not include AMS so the next line is not needed
% , and cosmic-ray antiproton signatures and 
by imposing that the signal is compatible with exclusion limits at 95\% CL.

\begin{figure}[h!]
\centering
\includegraphics[width=.49\textwidth]{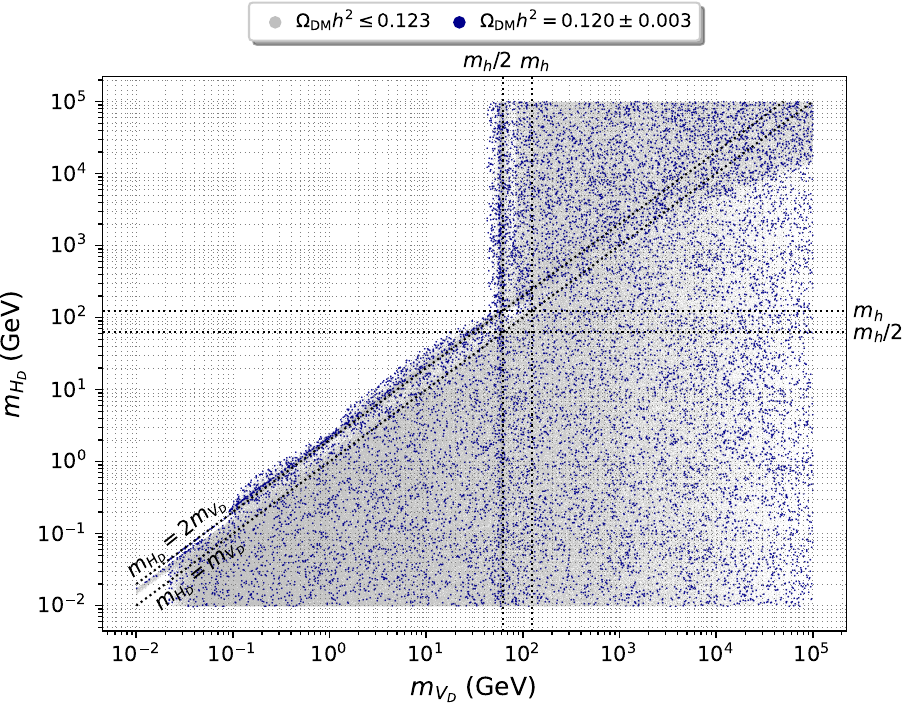}
\includegraphics[width=.49\textwidth]{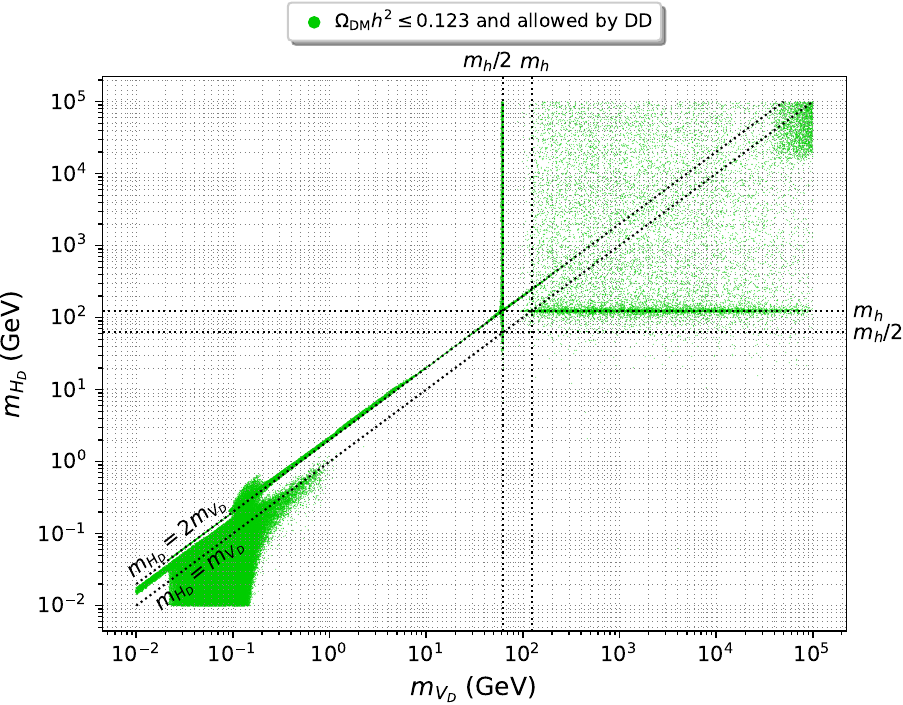}\\
\includegraphics[width=.49\textwidth]{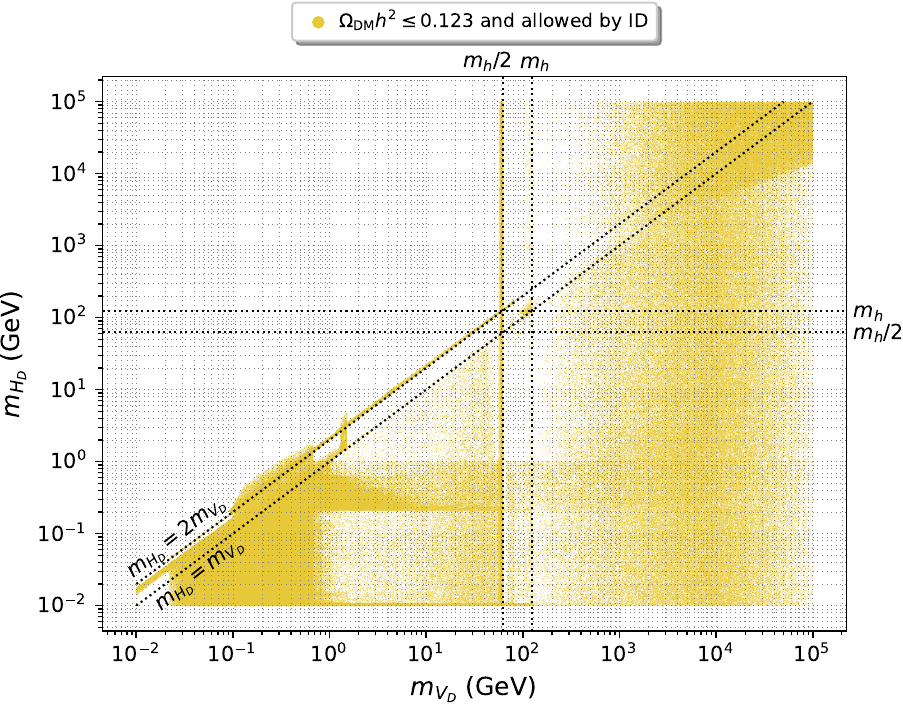}
\includegraphics[width=.49\textwidth]{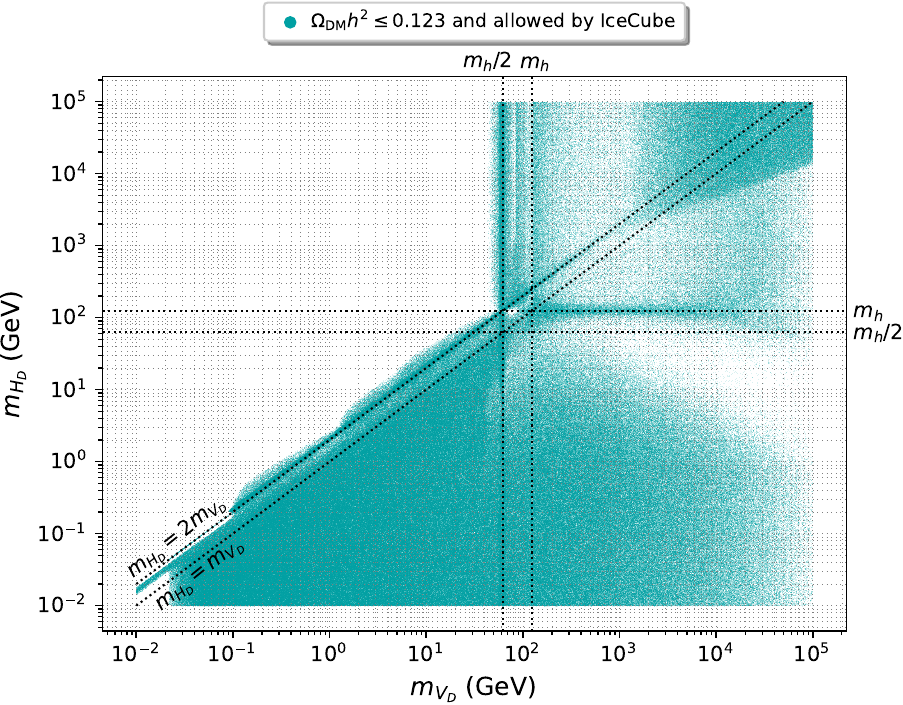}\\
\caption{\label{fig:cosmobounds} Areas in the $\{m_{V_D},m_{H_D}\}$ projection of the parameter space which are allowed by astrophysical and cosmological constraints: underabundant and exact relic density (top left), underabundant and allowed by direct detection experiments (top right), underabundant and allowed by indirect detection from Fermi-LAT (bottom left) and underabundant and allowed by neutrino fluxes measured at IceCube (bottom right).}
\end{figure}

All these constraints identify different allowed regions in the parameter space, shown in \cref{fig:cosmobounds} in the $\{m_{V_D},m_{H_D}\}$ projection, where in all panels we impose that relic density cannot be overabundant.
The region delimited by $m_{V_D}\lesssim m_h/2$ is therefore always excluded by the fact that it predicts a too high relic abundance for any choice of $g_D$ and $\theta_S$.

In the whole remaining parameter space, configurations of parameters which give a DM relic density compatible with the Planck observed value~\cite{Planck:2018vyg} within 3$\sigma$ can be found. A few regions tend to accumulate solutions in our uniform scan: the areas around $m_h=m_{H_D}$, $m_{V_D}=m_h/2$ and $m_{H_D}=2m_{V_D}$. All of these areas lead to enhancement in the annihilation or coannihilation cross sections due to interferences and resonance contributions. The constraints from DD, ID and IC exclude large patches of the parameter space, with the direct detection constraints being by far the strongest. The areas which are not excluded are in all cases the same which feature resonances and large interferences between processes, but also those for which the set of experiments for which data are available do not have sensitivity, such as the region for very small $m_{V_D}$ for the DD constraints. The higher density in the top-right corner of all panels is due to the fact that the region with high $V_D$ and $H_D$ masses leads to multiple contributions to the reduction of the DM relic density and all the considered experiments lose sensitivity in that region of the mass parameter space.\\ 

It is illustrative to describe in the following some of the main features which lead to patches of exclusions for the different observables.

Considering direct detection, for sufficiently small mixing angle $\theta_S$, the contribution from the diagram involving SM states in Figure~\ref{fig:annihilationtopologies} is suppressed. DM annihilation and DD cross section are thus uncorrelated in that limit. Then in principle, for small enough mixing angle, it should be possible to find points yielding the observed relic abundance while evading DD constraints for the whole region $m_{H_D}< m_{V_D}$, including the empty region $m_{H_D}< \min(m_{V_D},m_h/2)$ in the upper right panel of~\cref{fig:cosmobounds}~\cite{Arcadi:2016qoz}. However, for such a low mixing, the $H_D$ decay into SM particles is more and more suppressed and the relic abundance quickly increases above the measured experimental value. 
For the same reason, if we just consider the allowed points from DD, when $g_D$ becomes small, the cross section would be too small almost anywhere to be detected, with the exception of a region where the $V_D$ mass is between $\sim 10$ GeV and $m_h/2$, and $H_D$ lighter than $\sim 2$ GeV, where the sensitivity of LZ peaks~\cite{LZ:2022ufs} and the lower limit on our $g_D$ scan is still giving a large enough cross section which is excluded. However, by combining DD limits with the requirement of having underabundant relic density, all of the points in the light $H_D$ area of the plot become excluded because the coupling needed to achieve the correct relic density would give a too high cross section for direct detection, with the exception of the region with very light DM which is outside the sensitivity of all experiments, and for the region with high $V_D$ and $H_D$ masses (top-right area of the plot). 

For what concerns indirect detection, below $m_{V_D}\sim 500$ MeV the photons emitted by DM annihilation are too soft regardless of the channel, Fermi-LAT loses all sensitivity and all points are allowed. To understand the feature appearing when $m_{H_D}$ is between 200 MeV and 1 GeV is physical, we refer to the first panel, third row, of Figs. 3 and 4 in~\cite{Ambrogi:2018jqj}, where the photon spectrum for different annihilation channels is computed for two different DM masses. Below 200 MeV, the main decay channels are into up and down quarks and, more suppressed, into electrons; while the hadronic decays will lead to too soft photons because it is kinematically not possible to produce pions at tree level, the suppressed electron decays can produce enough photons only if the coupling strength is large enough, which explains the low number of allowed points in that region, especially for light DM masses. Above $m_{H_D}\sim 200$ MeV two channels open for $H_D$ decays: muons and strange quarks. The strange quarks will produce photons through hadronisation and pion production, but their energy is too low to be detected. For muons, the number of photons arising from this channel is much lower than other channels and, as the DM mass decreases, the sensitivity of Fermi-LAT also decreases. As $m_{H_D}$ increases various kinematic thresholds for hadron production are crossed, leading to a large number of photons and a larger exclusion rate.

Let us finally move to the constraints from IceCube. The sensitivity of the experiment is limited to the neutrino energy range 39 GeV to $4\times 10^5$ GeV~(see \cite{Belanger:2015hra} and references therein). Indeed, we have verified that below a DM mass of 39 GeV there is no exclusion from this observable. The regions with lower density of allowed points in the higher $m_{V_D}$ region are due to an interplay of effects which enter the neutrino flux at Earth, described by Eq.4 in ~\cite{Belanger:2015hra}: they involve the capture rate in the Sun, the annihilation cross section of the DM in the sun and the BRs of this process into SM final states (which depend on the mass of $H_D$); for example decays into $W^+W^-$, $ZZ$ and $t\bar{t}$, when allowed, produce neutrinos with high energy within the sensitivity reach of IceCube, and the masses of $V_D$ and $H_D$ determine the annihilation cross section, and therefore the number of produced neutrinos.\\

\begin{figure}[h!]
\centering
\includegraphics[width=.6\textwidth]{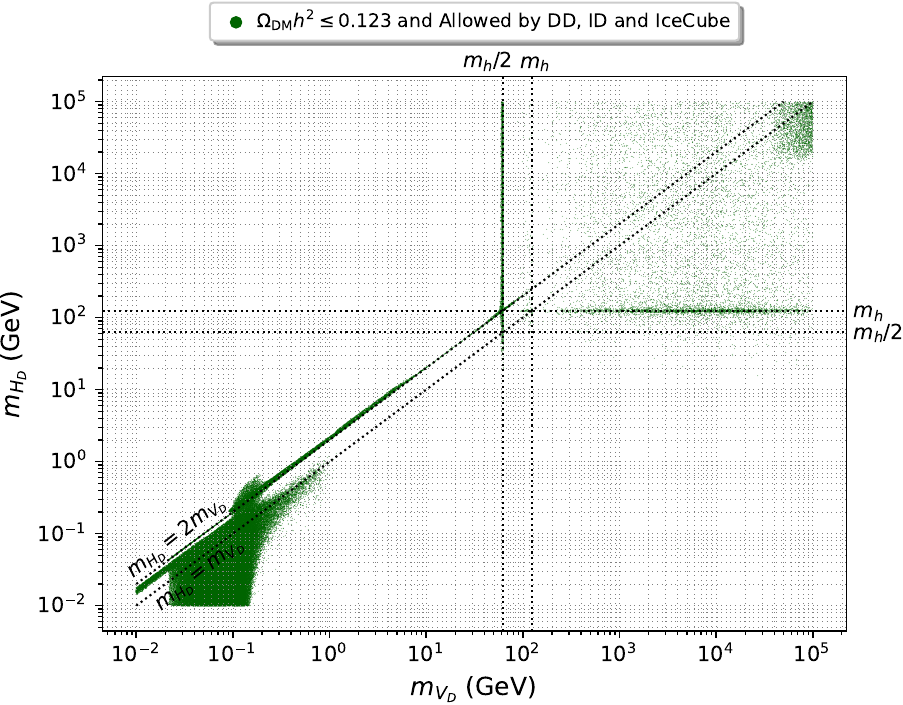}
\caption{\label{fig:combinationcosmobounds}Combination of all astrophysical and cosmological constraints in the $\{m_{V_D},m_{H_D}\}$ projection.}
\end{figure}

The combination of astrophysical and cosmological bounds, shown in \cref{fig:combinationcosmobounds}, leaves only the resonant areas, the area delimited by $m_{V_D}\gtrsim m_{h}$ and $m_{H_D}\gtrsim m_{h}/2$, and the region with $m_{V_D}\lesssim 100$ MeV where DD becomes insensitive. 
%It is interesting to see that even if there are regions allowed separately by the various observables, such as the area with $m_{H_D}\lesssim m_{h}/2$ and $m_{H_D}\lesssim m_{V_D}$, the values of the remaining parameters are complementary across the observables, such that the area is overall excluded when combining all of them.

\subsection{Theoretical and collider constraints}
\label{sec:thcollider}

% In any physical model, ensuring the stability of the vacuum is very important \cite{Benincasa:2022elt}. This means that the potential of the theory should be bounded from below, so that as the scalar fields grow large, the potential does not tend toward negative infinity. If the potential were to decrease indefinitely, it would indicate an unstable vacuum and would make the theory physically inconsistent. 
Requiring that the scalar potential is bounded from below leads to specific conditions on the couplings of the scalar fields defined in \Cref{eq:Lag}:
% The bounded-from-below (BFB) conditions ensure that the potential remains stable at high field values and are given by:
\begin{equation}
\lambda_H, \lambda_D \geq 0, \quad \lambda_{HD} \geq -2\sqrt{\lambda_H\lambda_D}\;
\end{equation}

% In addition to vacuum stability, another important requirement for the consistency of the theory is perturbative unitarity (PU). 
Perturbative unitarity (PU) requires that the scattering amplitudes for two-particle processes do not grow uncontrollably with energy. 
% In other words, the theory must stay well-defined and consistent at high energies, with no unphysical divergences from particle scattering. 
This imposes constraints on the combinations of coupling constants derived from the eigenvalues of the scattering matrix \cite{Branco:2011iw}:
 \begin{equation}
     \text{PU} = \left|(a_{+}, a_{-} , b_{+} , b_{-} , \lambda_{HD}) \right| < 8\pi,
 \end{equation}
where the parameters \(a_{+}\), \(a_{-}\), \(b_{+}\), and \(b_{-}\) are defined in terms of the couplings of the Higgs fields as follows:
\begin{equation}
\begin{aligned}
a_{\pm} &= 3 (\lambda_H + \lambda_D) \pm \sqrt{9 (\lambda_H - \lambda_D)^2 + 4 \lambda_{HD}^2}, \\
b_{\pm} &=  \lambda_H + \lambda_D \pm \left| \lambda_H - \lambda_D \right|.
\end{aligned}
\end{equation}
% The PU condition ensures that all scattering amplitudes remain within the physical limits imposed by the unitarity of the scattering matrix. In practice, this places a strict constraint on the values of the couplings \(\lambda_H\), \(\lambda_D\), and \(\lambda_{HD}\).

Electroweak precision observables (EWPO) serve as a valuable method for parametrising the impact of new physics on electroweak measurements. These observables are typically represented by the Peskin-Takeuchi parameters \(S\), \(T\), and \(U\) \cite{Peskin:1992sw,Peskin:1990zt}.
The contributions of our model to the \(S\), \(T\), and \(U\) parameters are \cite{Beniwal:2018hyi}
\begin{align}
 \label{eq:Tbar}
\overline{T} &= \frac{3}{16\pi s_W^2} \bigg[
\cos^2\theta_S \bigg( f_1\left(\frac{m_{h}^2}{m_W^2}\right) - \frac{1}{c_W^2} f_1\left(\frac{m_{h}^2}{m_Z^2}\right) \bigg) + \sin^2\theta_S \bigg( f_1\left(\frac{m_{H_D}^2}{m_W^2}\right)\nonumber \\
&\quad - \frac{1}{c_W^2} f_1\left(\frac{m_{H_D}^2}{m_Z^2}\right) \bigg) - \bigg( f_1\left(\frac{m_h^2}{m_W^2}\right) - \frac{1}{c_W^2} f_1\left(\frac{m_h^2}{m_Z^2}\right) \bigg)
\bigg],
\end{align}

\begin{align}
\label{eq:Sbar}
\overline{S} &= \frac{1}{2\pi} \bigg[
\cos^2\theta_S \, f_2\left(\frac{m_{h}^2}{m_Z^2}\right)
+ \sin^2\theta_S \, f_2\left(\frac{m_{H_D}^2}{m_Z^2}\right) 
- f_2\left(\frac{m_h^2}{m_Z^2}\right)
\bigg],
\end{align}

\begin{align}
\label{eq:Ubar}
\overline{U} &= \frac{1}{2\pi} \bigg[
\cos^2\theta_S \, f_2\left(\frac{m_{h}^2}{m_W^2}\right)
+ \sin^2\theta_S \, f_2\left(\frac{m_{H_D}^2}{m_W^2}\right) 
- f_2\left(\frac{m_h^2}{m_W^2}\right)
\bigg] - \overline{S},
\end{align}
where $f_1(x)$ and $f_2(x)$ are loop functions defined by \cite{Grimus:2008nb}:
\begin{align}
\label{eq:f1}
f_1(x) &= \frac{x \log x}{x - 1},
\end{align}

\begin{align}
\label{eq:f2}
f_2(x) &= 
\begin{cases} 
\frac{1}{12} \big[ -2x^2 + 9x + (x-3)(x^2 - 4x + 12) + \frac{1-x}{x} f_1(x) \\
\quad + 2\sqrt{(4-x)x} (x^2 - 4x + 12) \arctan\left(\sqrt{\frac{4-x}{x}}\right) \big], & 0 < x < 4, \\[1em]
\frac{1}{12} \big[ -2x^2 + 9x + (x-3)(x^2 - 4x + 12) + \frac{1-x}{x} f_1(x) \\
\quad + \sqrt{(x-4)x} (x^2 - 4x + 12) \log\left(\frac{x - \sqrt{(x-4)x}}{x + \sqrt{(x-4)x}}\right) \big], & x \geq 4.
\end{cases}
\end{align}
To constrain the parameter space of our model, the total values of \(S\), \(T\), and \(U\) must be within the confidence level of 95\% with respect to the measured ones \cite{ParticleDataGroup:2024cfk}. 

Finally, we checked the consistency of the parameters of our model by exploiting constraints from the Large Hadron Collider (LHC). This has been done with the {\tt HiggsTools} software ({\tt HiggsBounds} and {\tt HiggsSignals}) \cite{Bahl:2022igd}, which enforces bounds from direct searches for scalar particles, as well as the compatibility of the predictions of the Higgs coupling modification in our model with the corresponding measurements at the LHC.
The scalar mixing angle $\theta_S$ is strongly constrained by the signal strength measurement to $\vert \theta\vert \lesssim O(0.1)$, suppressing large deviations from SM-like behaviour, unless the new scalar is highly degenerate with the SM Higgs.

\begin{figure}[h!]
\centering
\includegraphics[width=.49\textwidth]{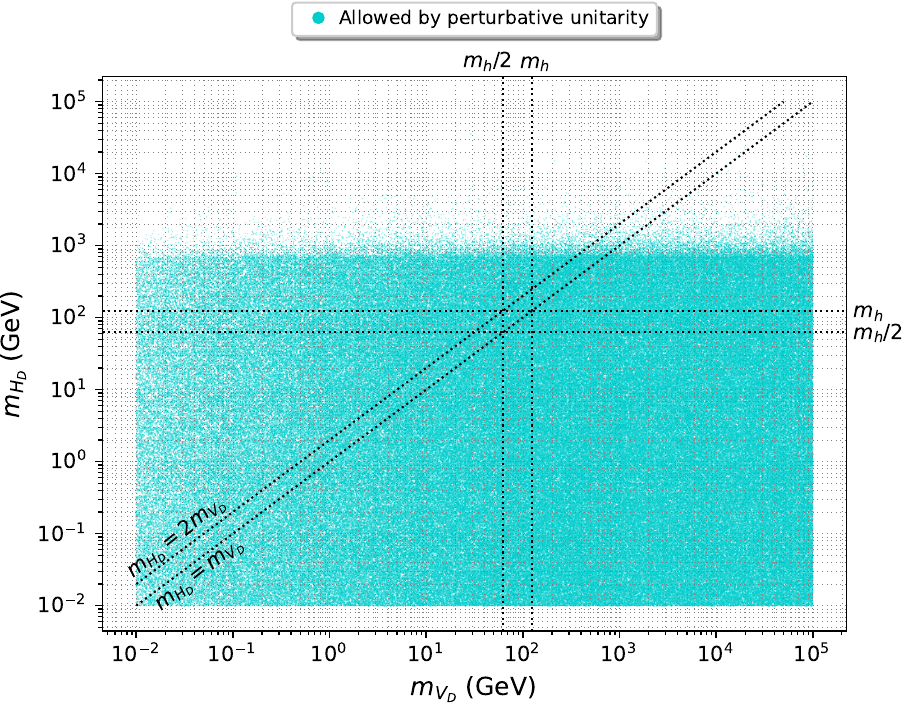}
\includegraphics[width=.49\textwidth]{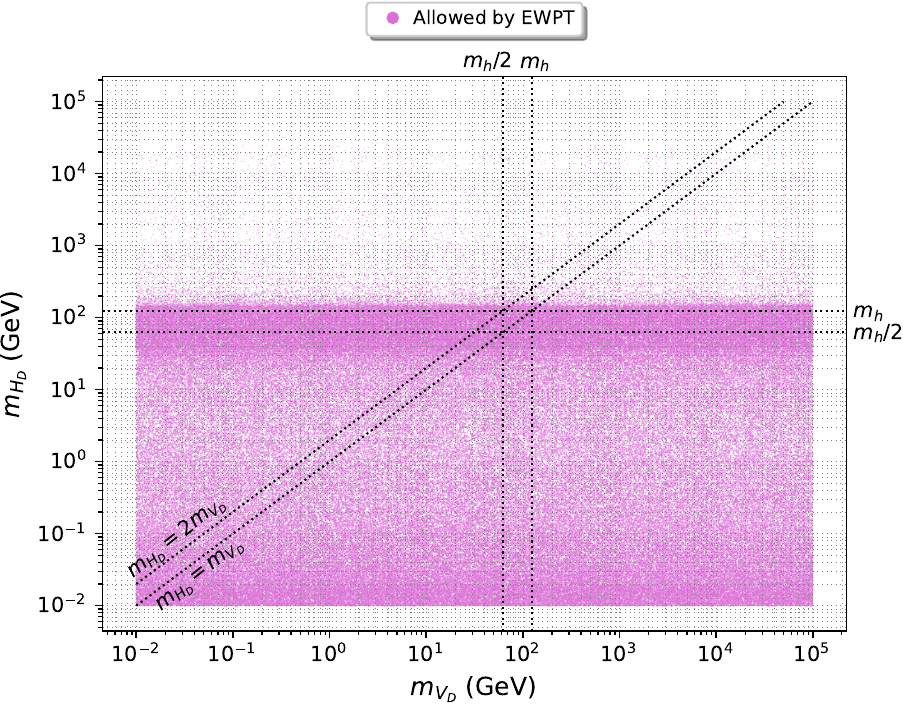}\\
\includegraphics[width=.49\textwidth]{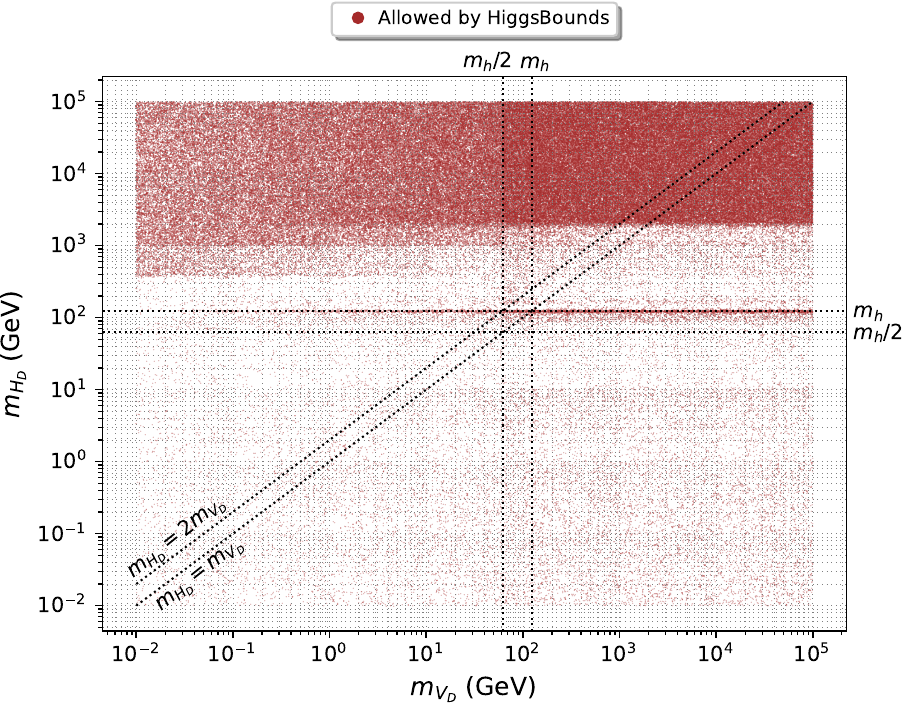}
\includegraphics[width=.49\textwidth]{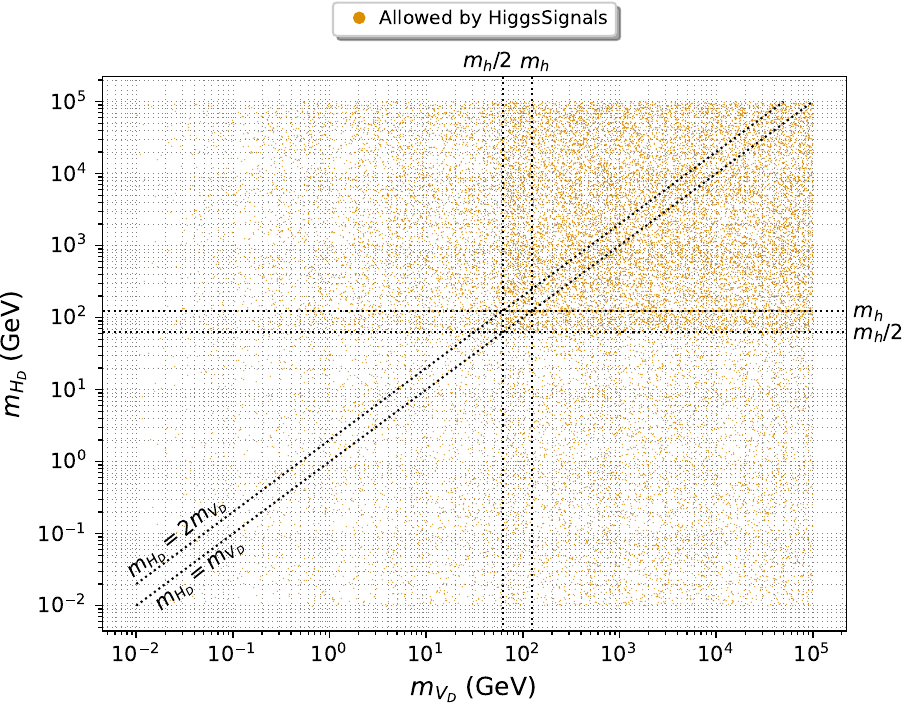}\\
\caption{\label{fig:collider} Allowed areas in the $\{m_{V_D},m_{H_D}\}$ projection of the $SU(2)_D$ model parameter space allowed by theoretical and collider constraints. From top left to bottom right: perturbative unitarity, EW precision tests, HiggsBounds and HiggsSignals.}
\end{figure}

\cref{fig:collider} displays the allowed regions in the $m_{V_D},m_{H_D}$ plane after imposing individual theoretical and collider constraints on our model parameter space, where one can see that PU and EWPO strongly limit the highest values achievable for the $H_D$ mass ($m_{H_D}\lesssim\mathcal O(10^4~\rm{GeV})$).\\

It is illustrative here to describe the horizontal bands appearing in the HiggsBounds results, which might look as artifacts, while in fact they represent well identified physical regions. These bands are horizontal because the dominant parameters for the determination of the collider constraints on the production and decay of the scalars of the theory are the $H_D$ mass and the scalar mixing angle, while the DM mass plays a subdominant role. Such bands are then related to the sensitivity regions of the various searches (both LEP and LHC) implemented in HiggsBounds and tend to exclude more strongly a lighter $H_D$, especially if the Higgs boson can decay into it, and allow heavier $H_D$ masses. The DM mass plays a more relevant role for searches looking for invisible decays, visible in this projection as different point densities for fixed $H_D$ mass, and depending on the search there can be sensitivity in the region where $m_{V_D}<m_h/2$, where the Higgs boson has a further invisible decay which can be excluded by dedicated LHC searches (see, {\it e.g.}~\cite{ATLAS:2018bnv,ATLAS:2023tkt}).\\

\begin{figure}[h!]
\centering
\includegraphics[width=.505\textwidth]{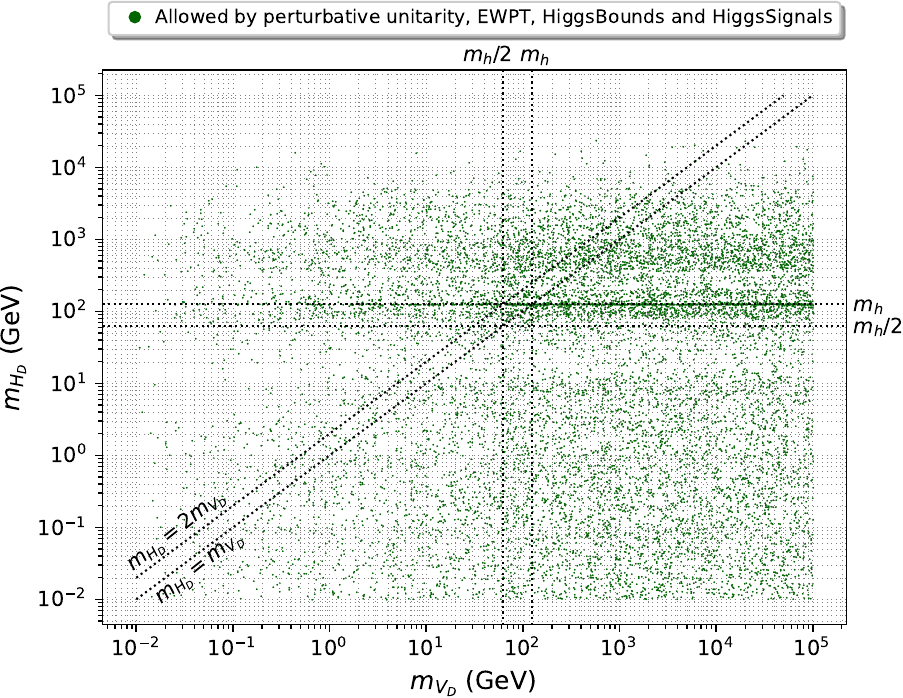}
\includegraphics[width=.47\textwidth]{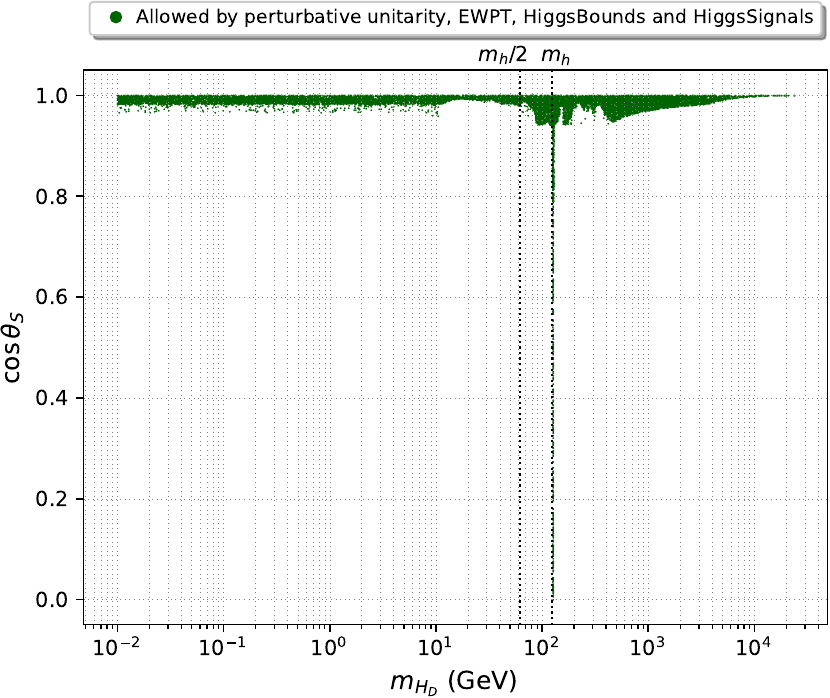}
\caption{\label{fig:combinationcollider}Combination of all collider and theoretical constraints in the $\{m_{V_D},m_{H_D}\}$ (left) and $\{m_{H_D},\cos\theta_S\}$ (right) projections.}
\end{figure}

% The top left panel shows the parameter space of the model that is allowed by the perturbative unitarity condition, whereas BFB is by construction automatically taken into account for any plots.   
\cref{fig:combinationcollider} presents the combined effect of all theoretical and collider constraints, showing the surviving parameter points in green on the $(m_{V_D}, m_{H_D})$ plane (left) and on the $(m_{H}, \cos\theta_S)$ plane (right). In the latter, it is clearly visible how the mixing angle is tightly constrained unless $H_D$ is almost degenerate with the Higgs boson, and the sensitivity areas of the aforementioned collider searches.

%\NB{Higgs signal strength measurements, $\vert \theta\vert \lesssim O(0.1)$}

\subsection{Phase transitions and gravitational waves}
\label{sec:PTGW}

\subsubsection{Loop corrections to scalar potential}

The tree-level potential is corrected at one-loop order in perturbation theory by the Coleman-Weinberg potential, which in the Landau gauge and in the $\overline{\text{MS}}$ renormalisation scheme reads as ~\cite{PhysRevD.7.1888}
\begin{equation}
    \label{eq:V1}
    V_1(\phi, \phi_D) = \frac{1}{64 \pi^2} \sum_i n_i m_i^4 \left( \ln \frac{m_i^2}{\mu^2} - C_i \right),
\end{equation}
where the sum ranges over the fields $i = W^\pm, Z, t, \phi, \phi_D, G^j, G_D^j, V_{D}$, $j\in\{0,+,-\}$, with $G$ and $G_D$ the SM and dark Goldstone bosons respectively (as usual, we retain only the top quark contribution and neglect all of the other lighter fermions). The numbers of bosonic and fermionic degrees of freedom are counted by $n_i$ and explicitly given by $n_{W^\pm} = 6, n_{Z} = 3, n_{V_D} = 9, n_t = -12, n_\phi = n_{\phi_D} = n_{G^j} = n_{G_D^j} = 1$. The constants $C_i$ are peculiar to the renormalisation scheme and in the $\overline{\text{MS}}$ 
are $C_i = 3/2$ for scalars, fermions, and $C_i = 5/6$ for vector bosons. Finally, the renormalisation scale can be chosen to be $\mu = v$. \\
The field-dependent masses in the scalar sector correspond to the eigenvalues of the mass matrix  
\begin{equation}
M^2_S =    \begin{pmatrix}
 -\mu_H^2+3\lambda_H \phi^2 + \frac{\lambda_{HD}}{2} \phi_D^2 &  \lambda_{\text{HD}}\phi\phi_D \\
  \lambda_{\text{HD}}\phi\phi_D & -\mu_D^2 + 3\lambda_D \phi_D^2 + \frac{\lambda_{\text{HD}}}{2}\phi^2
\end{pmatrix},
\end{equation} and to the Goldstone field-dependent masses,
\begin{equation}
    m^2_G = -\mu_H^2+\lambda_H \phi^2 + \frac{\lambda_{HD}}{2} \phi_D^2,\quad m^2_{G_D} = -\mu_D^2 + \lambda_D \phi_D^2 + \frac{\lambda_{\text{HD}}}{2}\phi^2,
\end{equation}
%\LDR{In $M_S^2$ the terms with $\mu^2, \mu_D^2$ are missing. Correct? We also need the equations for the field-dependent masses of the Goldstones}
while the field-dependent top quark and gauge boson masses are given by
\begin{equation}
m^2_t = \frac{y_t^2}{2}\phi^2,\quad m^2_{W^\pm}=\frac{g^2}{4}\phi^2,\quad m^2_{Z}=\frac{g'^2+g^2}{4}\phi^2,\quad m^2_{V_{D}} = \frac{g_D^2}{4}\phi_D^2,
\end{equation}
where $y_t$, $g'$, $g$ and $g_D$ are the top Yukawa coupling, the $U(1)_Y$, the $SU(2)_L$ and the $SU(2)_D$ gauge couplings, respectively. 
 
To change the renormalisation prescription one can add finite counterterms to the Coleman-Weinberg potential. Indeed, it is often convenient to directly use physical parameters as input values. For this purpose, one can enforce the VEVs, the scalar masses and the portal coupling to their tree-level values by including a counterterm potential 
\begin{equation}
    \label{eq:Vct}
    V_{\text{ct}}(\phi,\phi_D) = \delta\mu_H^2 \phi^2  + \delta\lambda_H \phi^4 +  \delta\mu_D^2 \phi_D^2 + \delta\lambda_D \phi_D^4 + \delta\lambda_{\text{HD}}\phi^2 \phi_D^2,
\end{equation}
in which the coefficients are fixed by the following renormalisation conditions
\begin{align}
\frac{\partial V_{\text{ct}}}{\partial \phi} \bigg|_\text{VEV} &= - \frac{\partial V_1}{\partial \phi} \bigg|_\text{VEV}, \\
\frac{\partial V_{\text{ct}}}{\partial \phi_D} \bigg|_\text{VEV} &= - \frac{\partial V_1}{\partial \phi_D} \bigg|_\text{VEV}, \\
\frac{\partial^2 V_{\text{ct}}}{\partial \phi^2} \bigg|_\text{VEV} &= - \left( \frac{\partial^2 V_1 |_{G^i = 0}}{\partial \phi^2} + \frac{1}{32 \pi^2} \sum_{i} \left( \frac{\partial m_{G^i}^2}{\partial \phi} \right)^2 \ln \frac{m^2_{\phi\text{IR}}}{\mu^2} \right) \bigg|_\text{VEV},  \\
\frac{\partial^2 V_{\text{ct}}}{\partial \phi_D^2} \bigg|_\text{VEV} &= - \left( \frac{\partial^2 V_1 |_{G^i_D = 0}}{\partial \phi_D^2} + \frac{1}{32 \pi^2} \sum_{i} \left( \frac{\partial m_{G_D^i}^2}{\partial \phi_D} \right)^2 \ln \frac{m^2_{\phi_D\text{IR}}}{\mu^2} \right) \bigg|_\text{VEV}, \\
\frac{\partial^2 V_{\text{ct}}}{\partial \phi \partial \phi_D} \bigg|_\text{VEV} &= - \frac{\partial^2 V_1}{\partial \phi \partial \phi_D} \bigg|_\text{VEV},
\end{align}
where VEV corresponds to the vacuum $(\phi, \phi_D) = (v, v_D)$ and the infrared regulators $m^2_{\phi\text{IR}}$ and $m^2_{\phi_D\text{IR}}$ prevent the singularities arising from the Goldstone boson contributions. Following \cite{Cline:2011mm} we adopt the choice $m^2_{\phi\text{IR}} = m_{h}^2$ and $m^2_{\phi_D\text{IR}} = m_{H_D}^2$.
The counterterms are explicitly given by
\begin{align}
    \delta\lambda_H &= \frac{1}{8 v^3} \left( \partial_\phi V_1 - \partial^2_{\phi^2}V_1 v \right)\bigg|_\text{VEV},\\
    \delta\lambda_D &= \frac{1}{8 v_D^3} \left( \partial_{\phi_D}V_1 - \partial^2_{\phi_D^2}V_1 v_D \right)\bigg|_\text{VEV},\\
    \delta\lambda_{HD} &= -\frac{\partial^2_{\phi\phi_D} V_1}{4 v v_D}\bigg|_\text{VEV},\\
    \delta\mu_H^2 &= \frac{1}{4 v} \left( \partial^2_{\phi^2} V_1 v - 3 \partial_\phi V_1 + \partial^2_{\phi\phi_D} V_1 v_D\right)\bigg|_\text{VEV},\\
    \delta\mu^2_D &= \frac{1}{4 v_{D}}\left(\partial^2_{\phi_D^2} V_1 v_D - 3 \partial_{\phi_D} V_1 + \partial^2_{\phi\phi_D} V_1 v\right)\bigg|_\text{VEV},
\end{align}
where the derivatives are evaluated at the tree-level vacuum.

The one-loop finite-temperature contributions to the effective potential are given by~\cite{Dolan:1973qd}
\begin{equation}
    \label{eq:VT}
    V_{\rm T} (\phi, \phi_D) = \frac{T^4}{2\pi} 
    \sum_i n_i J_{\text{B/F}}\left(\frac{m_i^2}{T^2}\right),
\end{equation}
where the thermal bosonic and fermionic functions $J_{\text{B/F}}$ are defined as
\begin{equation}
J_{\text{B/F}}(y^2) = \int_0^\infty dx~x^2 \ln\left(1\mp e^{-\sqrt{x^2+y^2}}\right).
\end{equation}
It is customary to deal with IR divergences arising from the zero bosonic Matsubara mode in the high-temperature limit by considering the so-called daisy resummation~\cite{RevModPhys.53.43}. This procedure amounts to adding the leading-order thermal correction or Debye mass $\Pi_i T^2$ to the field-dependent mass $m_i^2$ in $V_1$ and $V_T$~\cite{Parwani:1991gq}. For the scalar sector it consists in the shifts
\begin{align}
\mu_H^2 &\rightarrow \mu_H^2 + c_H T^2, \notag\\
\mu_D^2 &\rightarrow \mu_D^2 + c_D T^2,
\end{align}
where
\begin{align}
c_H &= \frac{1}{16}(g'^2 + 3 g^2 + 4 y_t^2) + \frac{\lambda_H}{2} + \frac{\lambda_{\text{HD}}}{6}, 
\\ 
c_D &= \frac{3}{16}g_D^2 + \frac{\lambda_D}{2} + \frac{\lambda_{\text{HD}}}{6},\label{eq:c2}
\end{align}
with the SM contribution taken from~\cite{PhysRevD.45.2933}. 

The thermal corrections also affect the longitudinal components of the gauge bosons. The mass of the $W_L$ is simply obtained by shifting $m_W^2(\phi)$ with the Debye mass $\Pi_{W_L} T^2 = \frac{11}{6} g^2 T^2$, while those of the neutral EW bosons, $Z_L$ and $\gamma_L$, are given by
\begin{align}
m_{Z_L}^2 &= \frac{1}{2}\left[ m_{Z}^2(\phi) + \frac{11}{6} \frac{g^2}{\cos^2 \theta_W} T^2 + \Delta \right], \\
m_{\gamma_L}^2 &= \frac{1}{2}\left[ m_{Z}^2(\phi) + \frac{11}{6} \frac{g^2}{\cos^2 \theta_W} T^2 - \Delta \right] 
\end{align}
with
\begin{equation}
\Delta^2 = \left(\frac{\phi^2}{4}+\frac{11}{6}T^2\right)^2(g'^2-g^2)^2+\frac{g'^2g^2}{4}\phi^4.
\end{equation}
These relations above are obtained from the diagonalisation of the thermally corrected mass matrix of the $(W_3, B)$ gauge fields. 
Finally, the contribution to the longitudinal part of the dark gauge boson $V_{D_L}$ is $\Pi_{V_{D_L}} T^2 = \frac{5}{6}g_D^2T^2$~\cite{Baldes:2018emh}.

\subsubsection{The phase transition}

A couple of parameters capture most of the information related to a first-order phase transition which proceeds through the nucleation of bubbles of true vacuum. The strength of the transition is characterized by the vacuum energy $\Delta \epsilon$ released during the nucleation process. The corresponding dimensionless parameter, $\alpha$, is obtained by normalising this vacuum energy to the energy density of the plasma in the symmetric phase $\rho_{\text{rad}}$~\cite{Espinosa:2010hh}:
\begin{equation}
\label{eq:alpha}
\alpha \equiv\frac{\Delta\epsilon}{\rho_{\text{rad}}}\Big|_{T=T_*},   \quad \Delta \epsilon \equiv \epsilon \big |_{\text{false vacuum}} - \epsilon \big |_{\text{true vacuum}}
\end{equation}
with
$\epsilon = V_{\rm eff} - \frac{T}{4}\frac{\partial V_{\rm eff}}{\partial T}$, $T_*$ the temperature at which GWs are generated ($T_* \simeq T_n$, where $T_n$ is the nucleation temperature) and $\rho_{\text{rad}} = \frac{\pi^2}{30} g_{\rm{eff}}(T) T^4$, where $g_{\rm{eff}}(T)$ is the effective number of relativistic degrees of freedom at the temperature $T$. 
%Note that a first-order phase transition is characterized as a strong one if the ratio of the VEV to the temperature, evaluated at $T_*$, satisfies
%\begin{equation}
%    \frac{v_*}{T_*}\geq 1.
%\end{equation}
%If this ratio is much smaller than unity the nature of the PT cannot be distinguished without lattice calculations, as also evidenced by the numerically very small value of the would-be GW signal in this case \cite{Kajantie:1996mn,Niemi:2020hto,Biekotter:2022kgf}.
The inverse time duration of the phase transition, $\beta$, is usually normalised by the Hubble parameter $H_*\equiv H(T_*)$~\cite{Grojean:2006bp}:
\begin{equation}
    \label{eq:beta}
    \frac{\beta}{H_*} = T\frac{d(S_3/T)}{dT}\Big |_{T=T_*},
\end{equation}
where $S_3$ is the three-dimensional Euclidean action computed for an $O(3)$-symmetric critical bubble. A large $\beta/H_*$ means that we can safely neglect the Hubble expansion of the Universe during the phase-transition process.

\subsubsection{The gravitational waves}

A single bubble cannot generate GWs, because its quadrupole moment vanishes due to its spherical symmetry. One of the key ingredients for a stochastic GW background  to be generated is the collisions between these bubbles of true vacuum which clearly break the aforementioned spherical symmetry~\cite{Kamionkowski:1993fg}. For bubbles evolving in a thermal plasma, additional contributions to the GW signal come from the metric perturbation induced by the perturbed plasma: sound waves~\cite{Hogan:1986qda, Hindmarsh:2013xza} and magnetohydrodynamic (MHD) turbulence~\cite{Caprini:2009yp}. In the former case, subsonic and supersonic bubbles, respectively, create compression waves beyond the bubble wall and rarefaction waves behind it. When acoustic waves from different bubbles overlap, the induced shear stress of the metric generates gravitational waves. Finally, bubble collision is also responsible for the turbulent motion of the fully ionized plasma and these MHD turbulences source a GW background. \\ The power spectrum $h^2\Omega_{\text{GW}}$ of the stochastic GW background from a cosmic phase transition thus arises from three contributions~\cite{Caprini:2015zlo}: $h^2\Omega_{\text{GW}}\simeq h^2\Omega_{\text{col}} + h^2\Omega_{\text{sw}} + h^2\Omega_{\text{turb}}$. %\footnote{In situations of strong supercooling, the GW power spectrum from sound-waves becomes similar to the contribution from bubble collision~\cite{Lewicki:2022pdb}.}
 In our model, we find that the dominant contribution is given by sound waves~\cite{Hindmarsh:2017gnf, Caprini:2019egz, Schmitz:2020rag}:

% The scalar-field contribution redshifted to today is given in the envelope approximation by~\cite{Huber:2008hg}\footnote{Recent works on contribution from bubble collisions can be found in~\cite{Lewicki:2019gmv, Lewicki:2020jiv, Lewicki:2020azd, Lewicki:2022pdb}.} 
% \begin{equation}
%  h^2\Omega_{\text{col}}(f) = h^2\Omega_{\text{col}}^{\text{peak}} S_\text{col}(f),
% \end{equation}
% with
% \begin{equation}
% \begin{split}
%   h^2\Omega_{\text{col}}^\text{peak} &= 1.67\times 10^{-5}\left(\frac{H_*}{\beta}\right)^2\left(\frac{\kappa_\text{col}\alpha}{1+\alpha}\right)^2\left(\frac{100}{g_*}\right)^{1/3}\left(\frac{0.11 v_w^3}{0.42+v_w^2}\right),  
%   \\
%    \quad S_\text{col} &= \frac{3.8\left(f/f_\text{col}\right)^{2.8}}{1+2.8\left(f/f_\text{col}\right)^{3.8}},
% \end{split}
% \end{equation}
% where $\kappa_\text{col}$ is the efficiency factor for the conversion of the vacuum energy into the gradient energy of the scalar field (energy stored in the shell of the scalar-field bubbles), $v_w$ is the bubble-wall speed in the rest frame of the plasma far away from the bubble~\cite{Caprini:2015zlo} and $f_\text{col}$ is the peak frequency, thus the frequency at $h^2\Omega_{\text{col}}^\text{peak}$.

%The sound-wave contribution is given by~\cite{Hindmarsh:2017gnf, Caprini:2019egz, Schmitz:2020rag}

\begin{equation}
 h^2\Omega_{\text{GW}}(f) \simeq h^2\Omega_{\text{sw}}(f) = h^2\Omega_{\text{sw}}^{\text{peak}} S_\text{sw}(f),
\end{equation}
with
\begin{equation}
\begin{split}
h^2\Omega_{\text{sw}}^\text{peak} &= 1.23\times 10^{-6}\left(\frac{H_*}{\beta}\right)\left(\frac{\kappa_\text{sw}\alpha}{1+\alpha}\right)^2\left(\frac{100}{g_*}\right)^{1/3}v_w\Upsilon, 
\\
S_\text{sw}(f) &= \left(\frac{f}{f_\text{sw}}\right)^3\left(\frac{7}{4+3\left(f/f_\text{sw}\right)^{2}}\right)^{7/2},
\end{split}
\end{equation}
where $\kappa_\text{sw}$ is the efficiency factor for the conversion of the vacuum energy into the bulk motion of the plasma and $f_\text{sw}$ is the sound-wave peak frequency\footnote{Note that the perturbative treatment of cosmic phase transitions may suffer from uncertainties due to gauge or renormalisation-scale dependence that could potentially lower the resulting GW signal stength~\cite{Croon:2020cgk, Athron:2022jyi}.}.
The suppression factor for a radiation-dominated Universe is defined as~\cite{Guo:2020grp, Ellis:2019oqb} 
\begin{equation}
\Upsilon=1-\frac{1}{\sqrt{2\tau_\text{sh}H + 1}},
\end{equation}
where $\tau_\text{sh}$ is the time after which sound waves do not source the GW production and where $\Upsilon \simeq \tau_\text{sh}H$ when $\tau_\text{sh}H \ll 1$, thus when this time is much smaller than a Hubble time.
%%%%%%%%%%%%%%%%%%%%%%%%%%%%%%%%%%%%%%%%%%%%%%%%%%%%%%%%%%

% Finally, the MHD-turbulence contribution is given by~\cite{Caprini:2015zlo, Schmitz:2020syl}
% \begin{equation}
%  h^2\Omega_{\text{turb}}(f) = h^2\Omega_{\text{turb}}^{\text{peak}} S_\text{turb}(f),
% \end{equation}
% with
% \begin{align}
% h^2\Omega_{\text{turb}}^\text{peak} &= 3.35\times 10^{-4}\left(\frac{H_*}{\beta}\right)\left(\frac{\kappa_\text{turb}\alpha}{1+\alpha}\right)^{3/2}\left(\frac{100}{g_*}\right)^{1/3}v_w\left[\frac{1}{2^{11/3}\left(1+8\pi f_\text{turb}/h_*\right)}\right], \\
% S_\text{turb} &= \frac{\left(f/f_\text{turb}\right)^3}{\left[1+\left(f/f_\text{turb}\right)\right]^{11/3}} \left[\frac{2^{11/3}\left(1+8\pi f_\text{turb}/h_*\right)}{1+8\pi f/h_*}\right],
% \end{align}
% where $\kappa_\text{turb}$ is the efficiency factor for the conversion of the vacuum energy into the MHD turbulences, $f_\text{turb}$ is the MHD-turbulence peak frequency and $h_*$ is the today-redshifted value of the Hubble rate at $T_*$.

\section{Combined results and discussion}
\label{sec:allresults}

\begin{figure}[h!]
\centering
\includegraphics[width=.7\textwidth]{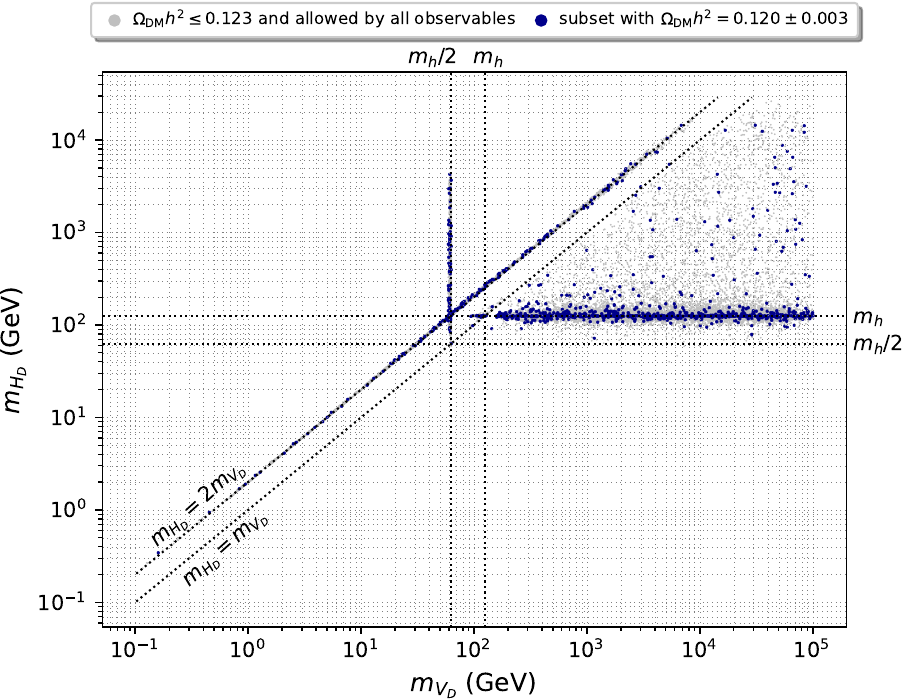}
\caption{Points in the $\{m_{V_D},m_{H_D}\}$ projection of the parameter space with underabundant relic density and satisfying all theoretical, collider and astrophysical constraints. Highlighted in dark blue colour are the points which reconstruct the measured relic density within 3$\sigma$. }
\label{fig:summaryscatter}
\end{figure}

The combination of all theoretical, cosmological, astrophysical and collider bounds is presented in the $\{m_{V_D},m_{H_D}\}$ projection of the parameter space in \cref{fig:summaryscatter}, where we show both points which lead to an underabundant DM relic density (meaning that other DM components must exist) and those for which the gauge boson DM of the $SU(2)_D$ model is sufficient to explain the entire observed abundance. Notice that the difference in the density of allowed but just underabundant points and allowed points with exact relic density is just meant to represent graphically the proportion of the latter with respect to the former. Indeed, by (largely) increasing the number of scanned points, it would be possible to fill the allowed areas with points with exact relic density, but this would be computationally extremely expensive and beyond the scope of this analysis.

The distribution of points precisely identifies specific regions of the parameter space: they span a range delimited by $m_h/2 \lesssim m_{H_D} \lesssim m_{V_D}$ and cluster on three narrow regions characterised by $m_{H_D}\simeq m_h$ or $m_{V_D}\simeq m_h/2$ or $m_{H_D}\simeq 2m_{V_D}$. The $m_{V_D}\simeq m_h/2$ or $m_{H_D}\simeq 2m_{V_D}$ values make the DM annihilation processes with a resonant scalar in the $s$-channel effective, thus reducing the relic density even for small $g_D$ (within the scanned range). When the two scalars are almost degenerate, on the other hand, their mixing can be large, and therefore both scalars can decay to the same channels as the SM Higgs boson.

\begin{figure}[h!]
\centering
\includegraphics[width=.99\textwidth]{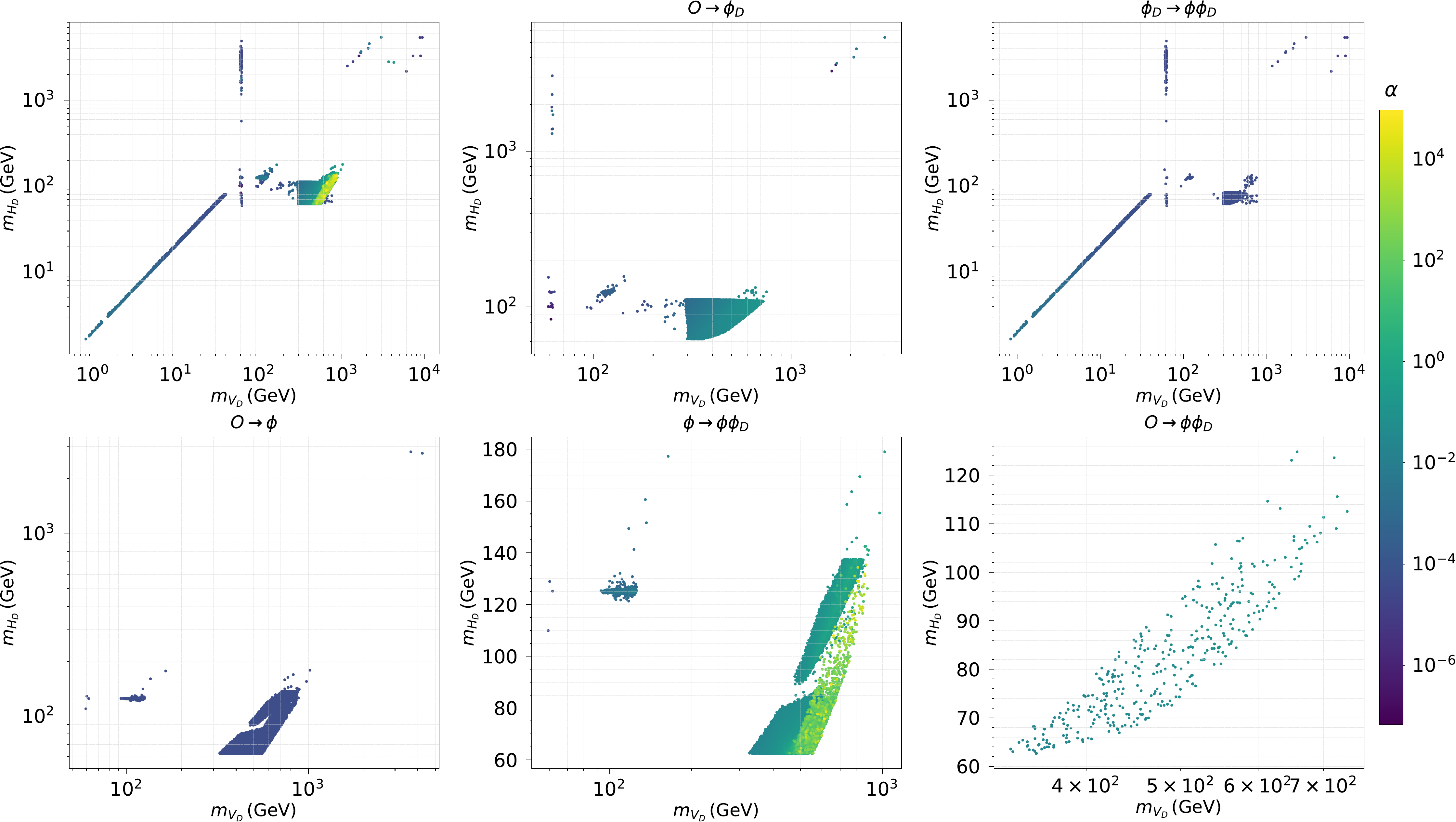}
\caption{Projection of the parameter space on the plane spanned by $m_{V_D}$ and $m_{H_D}$. The upper-left panel shows the full results of our scan, while the other panels only show the results for a specific PT step. The colour code, for each panel, represents the PT strength $\alpha$.}
\label{fig:plot1}
\end{figure}

% \LP{Write a sentence about filtering the points for gD less than 1, to be conservative}
Next, we analyse the phase transition dynamics for each point in Figure~\ref{fig:summaryscatter} (in addition to a refined scan, see below) %of the considered parameter space
via the \texttt{CosmoTransitions} package~\cite{Wainwright:2011kj}. Our results are shown in Figure~\ref{fig:plot1}, which is a projection of the scanned parameter space. It indicates which values of the DM mass $m_{V_D}$ and the dark Higgs mass $m_{H_D}$ lead to a strong FOPT. We can identify three main regions, as in Figure~\ref{fig:summaryscatter}. One corresponds to $m_{V_D}\simeq m_h/2$, while another one is identified by the condition $m_{V_D}\simeq  m_{H_D}/2$. Points with symmetry non-restoration are found in the latter region: the minimum of the potential is at $(0,v_D')$ at high temperature, instead of $(0,0)$ because the coefficient $c_D$ in Eq.~(\ref{eq:c2}) is negative, therefore preventing the symmetry to be restored as the temperature increases. Finally, the last region is broader and corresponds to $m_{V_D}\in [4\times10^2, 10^3]$ GeV, with $\vert m_{H_D}-m_h\vert\lesssim m_h/2$ GeV. We find that the strong FOPTs are obtained in this region and that some of these points lead to the observed DM relic density. We then make a refined scan in this strong FOPT region to find more such points. The combined results of the general and refined scan are shown in the upper-left panel in Figure~\ref{fig:plot1}. We also show each PT step contribution in the remaining panels. We then immediately see that the strongest phase transitions are obtained for the PT step $\phi\rightarrow \phi\phi_D$, from the double-step FOPT $O\rightarrow\phi\rightarrow \phi\phi_D$, where in this notation $O$ corresponds to the origin $(0,0)$, $\phi$ corresponds to a phase along the $\phi$ axis, $(v',0)$ and $\phi\phi_D$ corresponds to a phase in the plane, $(v'',v_D'')$. The reason why, although this strong FOPT region contains points with different PT patterns ($O\rightarrow\phi_D\rightarrow\phi\phi_D$, $O\rightarrow\phi\rightarrow\phi\phi_D$ and $O\rightarrow\phi\phi_D$, only the second one yields $\alpha \geq 1$), is the following: This is because, as in~\cite{Benincasa:2022elt}, PTs starting from the origin are found to be weaker. Moreover, the transition $\phi_D\rightarrow\phi\phi_D$ cannot be arbitrarily strong as the value of the VEV along the $\phi$ axis is bounded by $v\simeq 246$ GeV. However, for the transition $\phi\rightarrow\phi\phi_D$, a large $v_D(T_n)$ necessarily leads to a large $\alpha$, as the latter is proportional to $v_D(T_n)/T_n$, with $T_n\sim 10^2$ GeV. Indeed, these strong FOPTs come from points with $v_D(T=0)\in[10^3,2\times 10^3]$ GeV.  Finally, note that the two points in the upper-right corner of the lower-left panel give rise to a PT step along the direction $\phi\rightarrow\phi_D$. Actually, the full pattern for these two points is a triple-step FOPT $O\rightarrow \phi\rightarrow\phi_D\rightarrow\phi\phi_D$, from which the PT step $\phi\rightarrow\phi_D$ yields a strong enough GW to be detected. However, these two points lead to an underabundant DM, $h^2\Omega_\text{DM}\sim 0.01$.

\begin{figure}[h!]
\centering
\includegraphics[width=.99\textwidth]{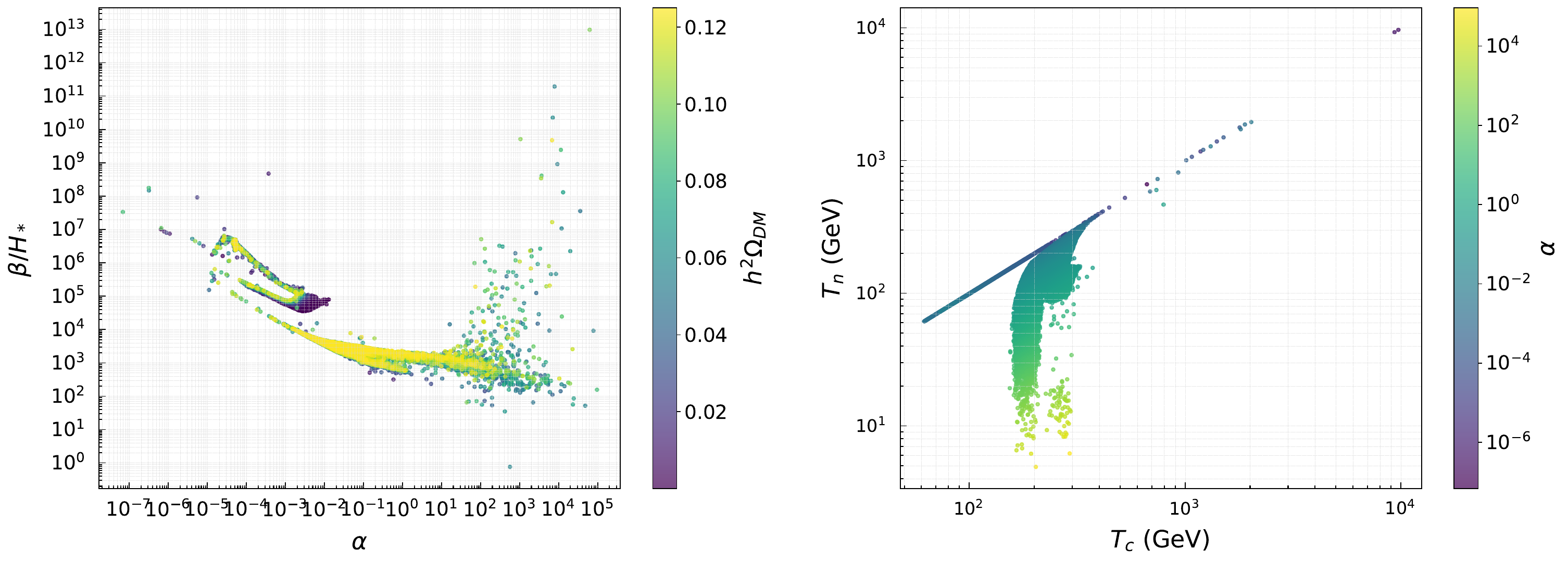}
\caption{The left panel shows the correlation between the PT strength $\alpha$ and the normalised inverse PT duration $\beta/H_*$. The colour bar indicates the DM abundance. The right panel shows the correlation between the critical temperature $T_c$ and the nucleation one $T_n$, with the PT strength $\alpha$ for the colour code.}
\label{fig:plot2}
\end{figure}

Next, in the left panel in \cref{fig:plot2}, we show the correlation between the phase-transition parameters $\alpha$ and $\beta/H_*$. As expected, we obtain that the slower the phase transition (smaller $\beta/H_*$), the stronger (larger $\alpha$) it is. Moreover, we find that in the present model, the DM abundance is not correlated to the PT strength, as we obtain the correct DM relic density for a large range of values of $\alpha$. In the right panel, the relevant temperatures for the study of phase transitions, namely $T_c$ and $T_n$, are displayed, with the colour code indicating the PT strength $\alpha$. Along the oblique line, we have $T_n \simeq T_c$, which means that the PT occurs as soon as the new minimum in the potential becomes more stable than the current phase. It therefore results in a small value of $\alpha$. By contrast, the more $T_n$ deviates from the line $T_n = T_c$, the slower is the phase transition, and as shown in the left panel, the bigger the PT strength $\alpha$. This shows that $\alpha$ and the ratio $T_n/T_c$ are negatively correlated: the smaller $T_n/T_c$ (larger $\alpha$), the larger the amount of supercooling.

\begin{figure}[h!]
\centering
\includegraphics[width=.99\textwidth]{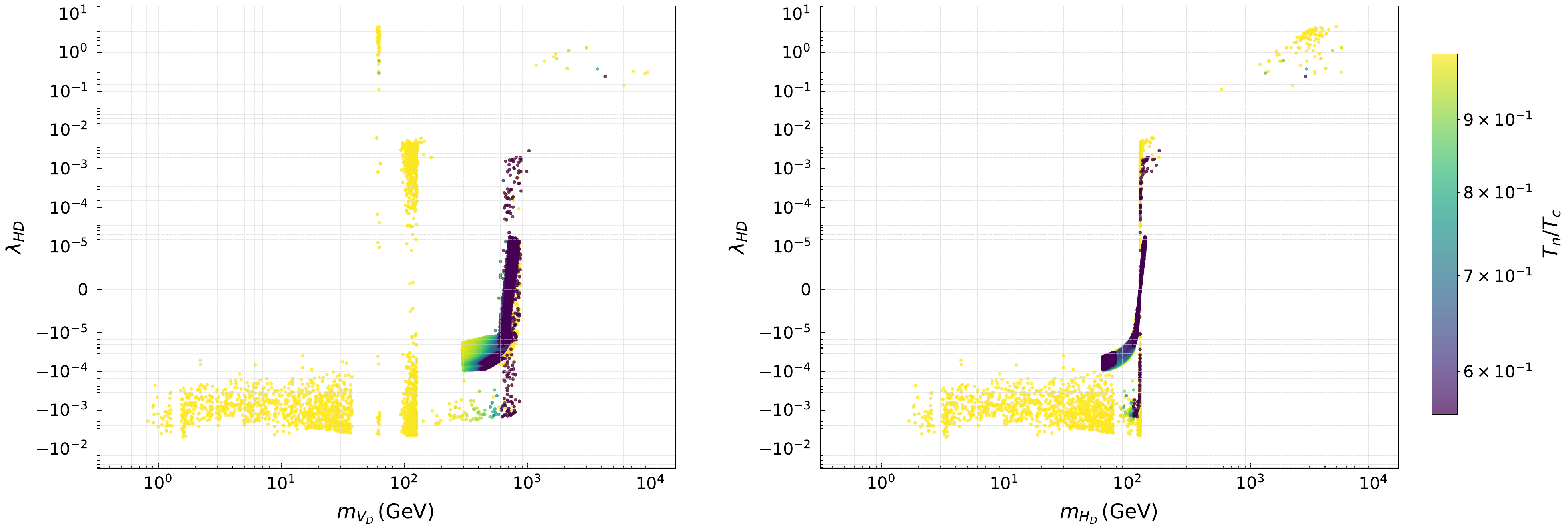}
\caption{The left panel shows the correlation between the portal coupling $\lambda_{\rm{HD}}$ and the DM mass $m_{V_D}$, while the right panel shows the correlation between $\lambda_{\rm{HD}}$ and the dark Higgs mass $m_{H_D}$. For both panels, the colour bar shows the value of the ratio $T_n/T_c$.}
\label{fig:plot3}
\end{figure}

Then, in \cref{fig:plot3}, we show how the portal coupling $\lambda_{\rm{HD}}$ is correlated to the masses $m_{V_D}$ and $m_{H_D}$. We can first notice that in most of the parameter space leading to FOPTs, $\vert\lambda_{\rm{HD}}\vert$ is small: $\vert\lambda_{\rm{HD}}\vert\lesssim 6\times10^{-3}$, and we have $m_{H_D}\lesssim 155$ GeV. Furthermore, a stronger FOPT, thus a smaller value of $T_n/T_c$, is preferred for $m_{V_D}\in [4\times10^2, 10^3]$ GeV and a moderate mass splitting $\vert m_{H_D}-m_h\vert\lesssim m_h/2$ GeV, as already stated when describing Figure~\ref{fig:plot1}. The refined scan yields points in the hockey-stick region bounded by $\lambda_\text{HD}\in[-10^{-4}, 2\times10^{-5}]$, where a lot of them both provide the observed DM relic density, as well as a strong enough FOPT to be detectable. In addition, we find that such strong FOPTs require $g_D\in [0.6, 1]$.

% \begin{figure}[h!]
% \centering
% \includegraphics[width=.7\textwidth]{./Figures/plot4.png}
% \caption{\label{fig:plot4}}
% \end{figure}
 
\begin{figure}[h!]
\centering
\includegraphics[width=.7\textwidth]{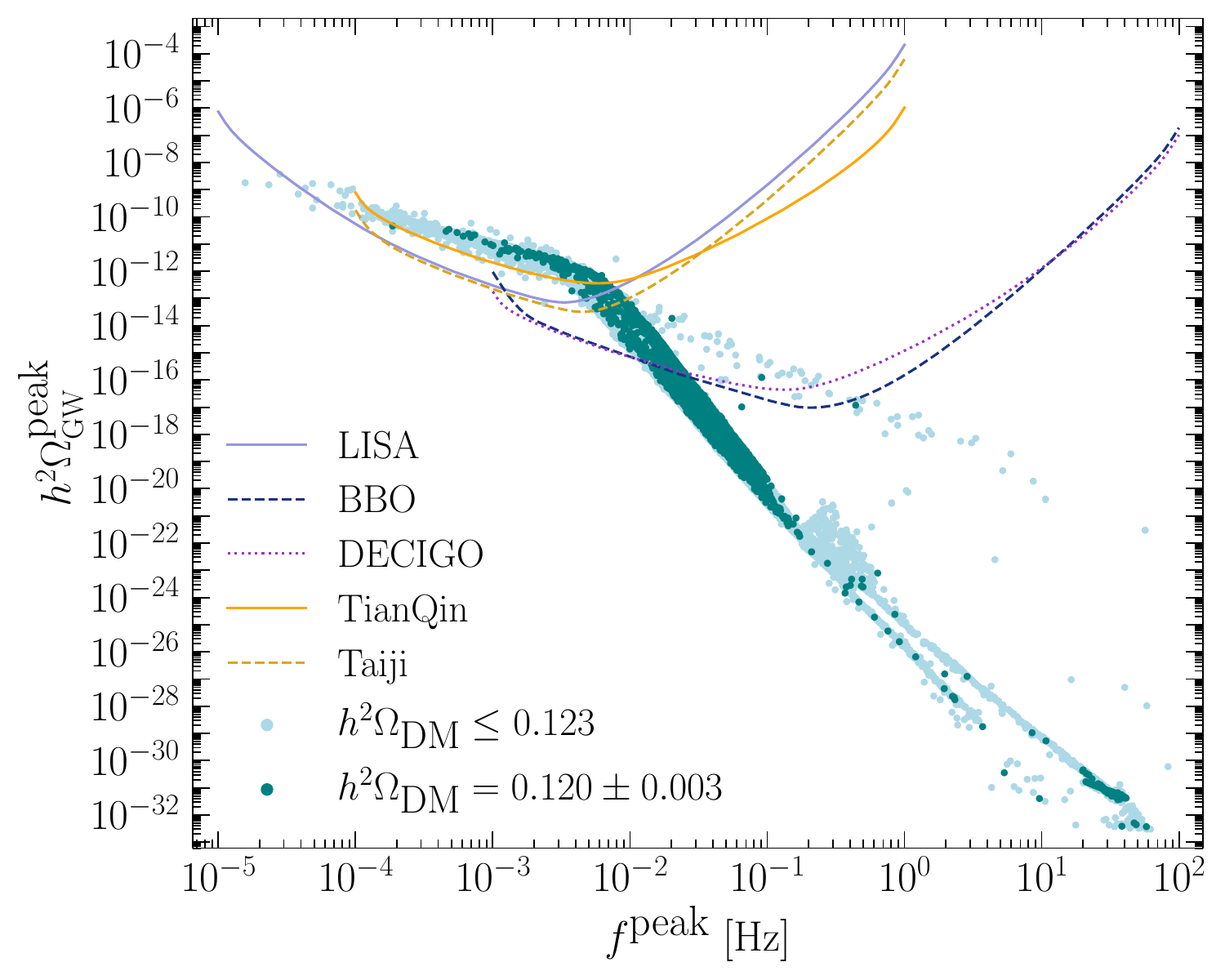}
\caption{Peak of the GW signal $h^2\Omega^{\text{peak}}_\text{GW}$ as a function of the peak frequency $f^\text{peak}$. Darker points are those yielding the observed DM relic density. The power-law integrated sensitivity curves of LISA, DECIGO, BBO, TianQin and Taiji are also shown.}
\label{fig:GW}
\end{figure}

\cref{fig:GW} shows the peak of the GW power spectrum $h^2\Omega^{\text{peak}}_\text{GW}$ from FOPTs and the corresponding frequency $f^\text{peak}$ of this spectrum at its peak. The darker points indicate those giving the correct DM abundance. The power-law-integrated sensitivity curves~\cite{Thrane:2013oya} of LISA, DECIGO, BBO, TianQin and Taiji for an observation time of 4 years and signal-to-noise ratio equal to 10, are displayed and we see that a significant number of points can potentially be detected by these five GW detectors. Moreover, a large part of these points also provide a DM abundance that matches the Planck observations. These dark-color points are essentially from the refined scan, which focuses on strong FOPTs. This is why points with a weaker GW signal, $h^2\Omega^{\text{peak}}_\text{GW} \lesssim 10^{-22}$, arising from the more general scan, are less populated than the dark points; the refined scan mostly provides points with $h^2\Omega^{\text{peak}}_\text{GW} \gtrsim 10^{-22}$.

\begin{figure}[h!]
\centering
\includegraphics[width=.49\textwidth]{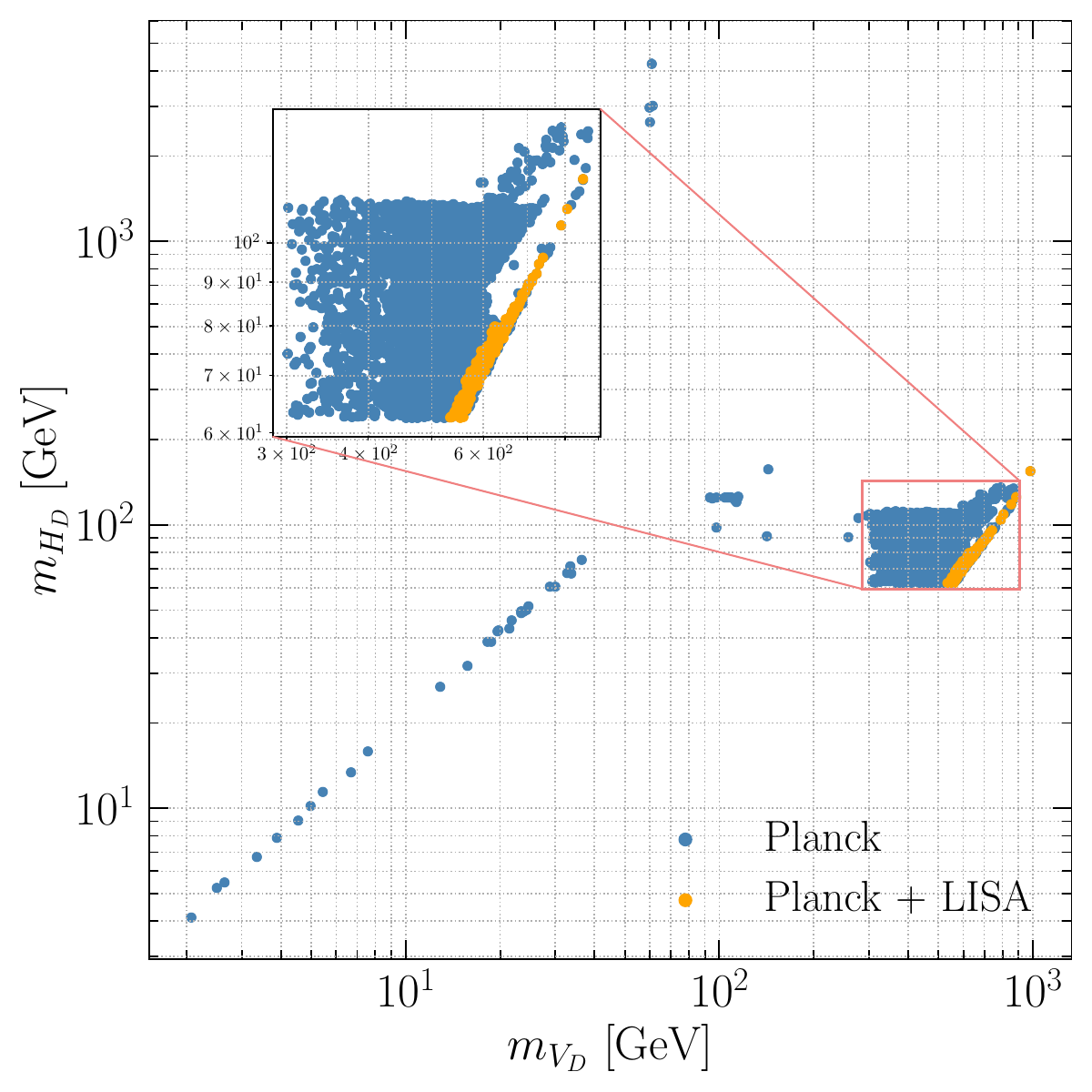}~~~
\includegraphics[width=.49\textwidth]{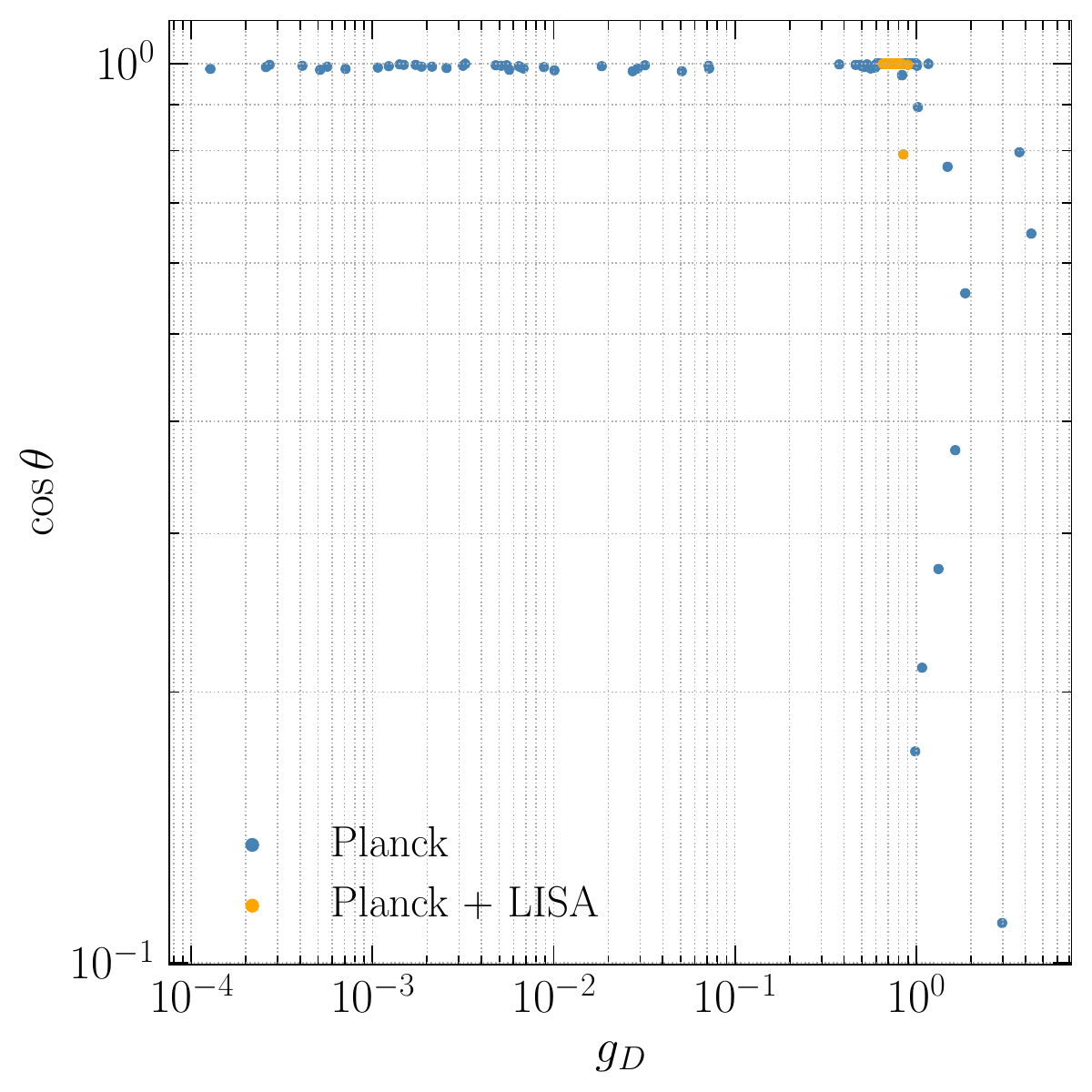}\\
\caption{\label{fig:planck-lisa} Results from both the general and refined scans indicating which values of the four free parameters lead to the observed DM relic density (blue) while also being detectable by LISA (orange). The enlarged box in the left panel only contains points from the refined scan. The left panel shows a projection on the plane spanned by the dark-sector masses $m_{V_D}$ and $m_{H_D}$, while the right panel shows the correlation between the remaining free parameters: $g_D$ and $\cos\theta$.}
\end{figure}

Finally, in \cref{fig:planck-lisa}, we show the values of the four free parameters of this model that can account for the observed DM relic density, as constrained by the Planck mission. These are depicted by blue points. Those points which, in addition, lead to a GW signal reaching the sensitivity of LISA are depicted in orange. As in \cref{fig:plot1}-\ref{fig:GW}, \cref{fig:planck-lisa} presents results from both the general and refined scans combined. The latter has been identified when describing \cref{fig:plot1} and correspond to a region of the parameter space leading to strong FOPTs. In order to clearly highlight points from this refined scan, we add an enlarged box containing only points from the latter in the left panel of \cref{fig:planck-lisa}. Regarding the right panel, points from the refined scan lie in the range $g_D\in[0.6,1]$, with $\cos\theta\simeq1$. We can then clearly see that the most interesting points, the orange ones, mostly come from this refined scan. They are approximately obtained for $m_{V_D}\in [500, 900]$ GeV, $m_{H_D}\in [m_h/2, 125 \text{~GeV}]$, $g_D\in [0.65, 0.85]$ GeV and $\cos\theta\simeq 1$.

\section{Conclusions}

In this paper we have discussed the possibility to generate gravitational waves during a first-order phase transition in a scenario where the Standard Model is extended by a new $SU(2)$ gauge group. In this scenario, none of the SM particles transform under this $SU(2)$, but the SM Higgs doublet can interact with the scalar doublet which generates the new gauge-boson masses through a quadrilinear term in the scalar potential. This also induces a mixing between the Higgs boson and the physical scalar degree of freedom which is left after spontaneous symmetry breaking. The custodial symmetry of the scalar sector noticeably implies that the new massive gauge bosons are stable, and therefore are dark matter candidates. 
We have tested this model against the most recent experimental data from collider and astrophysical observables and determined the allowed regions of its four-dimensional parameter space. 
In these regions, we found we could obtain the observed DM abundance and that strong first-order phase transitions are possible. We have then shown that the resulting power spectrum of the generated stochastic gravitational-wave background is well within the sensitivity range of different gravitational-wave detectors such as LISA, DECIGO, BBO, TianQin or Taiji.
We have thus shown how gravitational waves can provide a precious complementary way to probe scenarios of new physics, enhancing their discovery potential or at least narrow down their parameter space.

\section*{Acknowledgments}
N. B. is supported by the National Natural Science Foundation of China (Grant No. 12475105).
The work of L. P. is supported by ICSC – Centro Nazionale di Ricerca in High Performance Computing, Big Data and Quantum Computing, funded by the European Union – NextGenerationEU.
The work of L. D. R.
has been funded by the European Union – Next Generation EU through Research Grant No. P2022Z4P4B ``SOPHYA - Sustainable Optimised PHYsics Algorithms: fundamental
physics to build an advanced society'' under the program PRIN 2022 PNRR of the Italian
Ministero dell’Università e Ricerca (MUR) and by Research Grant No. 20227S3M3B
``Bubble Dynamics in Cosmological Phase Transitions'' under the program PRIN 2022 of
the Italian Ministero dell’Università e Ricerca (MUR).

\bibliographystyle{JHEP}
\bibliography{biblio}

\end{document}